\documentclass[%
 reprint,
 amsmath,amssymb,
 aps,nofootinbib,
]{revtex4-2}
\usepackage{tikz}
\usepackage{xcolor}
\usepackage{multirow}
\usepackage{graphicx}
\usepackage{dcolumn}
\usepackage{bm}
\usepackage{fancyhdr}
\usepackage[colorlinks]{hyperref}
\usepackage[nameinlink,noabbrev]{cleveref}
\hypersetup{
    colorlinks=true,
    linkcolor=blue,
    filecolor=magenta,      
    urlcolor=blue,
   citecolor=blue
}
\usepackage{graphicx}
\usepackage{dcolumn}
\usepackage{bm}
\usepackage{scalerel}
\usepackage{tikz}
\usetikzlibrary{svg.path}

\definecolor{orcidlogocol}{HTML}{A6CE39}
\tikzset{
  orcidlogo/.pic={
    \fill[orcidlogocol] svg{M256,128c0,70.7-57.3,128-128,128C57.3,256,0,198.7,0,128C0,57.3,57.3,0,128,0C198.7,0,256,57.3,256,128z};
    \fill[white] svg{M86.3,186.2H70.9V79.1h15.4v48.4V186.2z}
                 svg{M108.9,79.1h41.6c39.6,0,57,28.3,57,53.6c0,27.5-21.5,53.6-56.8,53.6h-41.8V79.1z M124.3,172.4h24.5c34.9,0,42.9-26.5,42.9-39.7c0-21.5-13.7-39.7-43.7-39.7h-23.7V172.4z}
                 svg{M88.7,56.8c0,5.5-4.5,10.1-10.1,10.1c-5.6,0-10.1-4.6-10.1-10.1c0-5.6,4.5-10.1,10.1-10.1C84.2,46.7,88.7,51.3,88.7,56.8z};
  }
}

\newcommand\orcidicon[1]{\href{https://orcid.org/#1}{\mbox{\scalerel*{
\begin{tikzpicture}[yscale=-1,transform shape]
\pic{orcidlogo};
\end{tikzpicture}
}{|}}}}
\usepackage{amsmath}
\usepackage{amssymb, amsfonts}
\usepackage[greek,english]{babel}

\begin{document}

\preprint{APS/123-QED}

\title{Beyond Maxwell-Boltzmann: Transport in Quasiequilibrium Plasmas}


\author{Kamel Ourabah \orcidicon{0000-0003-0515-6728},}\email{kam.ourabah@gmail.com, kourabah@usthb.dz}
\address{Theoretical Physics Laboratory, Faculty of Physics, University of Bab-Ezzouar, USTHB, Boite Postale 32, El Alia, Algiers 16111, Algeria}

\date{\today}

\begin{abstract}
Space plasmas are generally characterized by non-Maxwellian distributions with suprathermal populations, as routinely revealed by \textit{in situ} observations. Such departures from standard Maxwellian distributions can be understood as signatures of quasiequilibrium states, in which the distribution of the medium can be expressed as a continuous superposition of Maxwellian distributions, namely through superstatistics. Here, we construct macroscopic relations linking fluxes to their associated driving forces in such plasmas, where superstatistical effects enter the picture through the transport coefficients. After comparing the resulting superstatistical distributions with observed electron distributions in the solar wind, we turn to the kinetic response of quasiequilibrium plasmas and derive the corresponding transport coefficients, including the electric and thermal conductivities, the mobility, and the diffusion coefficient. We further extend the analysis to viscous plasmas and compute the shear and bulk viscosity coefficients. Overall, quasiequilibrium effects are found to systematically enhance the transport coefficients relative to their Maxwellian values. We quantify this enhancement for the three main universality classes of superstatistics, which are the most commonly encountered in experimental and observational situations, and interpret it as a consequence of the increased population of energetic particles in the non-Maxwellian tails.
\end{abstract}

\maketitle


\section{Introduction}
This year marks the 120th anniversary of the death of Ludwig Boltzmann, whose profound ideas laid the very foundations of statistical mechanics. What have we learned over these 120 years? While Boltzmann, or Maxwell–Boltzmann (MB), distributions are indeed observed in many situations, the accumulation of experimental and observational evidence has made it increasingly clear that departures from this classical behavior are far more common than one might have anticipated. Indeed, non-MB distributions now appear to be ubiquitous across a wide range of systems, at almost all scales, from controlled laboratory experiments \cite{cold1,cold2,HE1,HE2} to space \cite{azerty1,azerty2,azerty3,azerty4,azerty5} and astrophysical environments \cite{Astro1,Astro3,galaxy,Astro4}.

Perhaps the area where the strongest evidence for non-MB distributions has emerged, and indeed historically the domain where such deviations were first identified \cite{azerty1,azerty2}, is plasma physics. In particular, in space plasmas, non-MB distributions are the rule rather than the exception: the observed distributions are typically characterized by pronounced high-energy tails that decay more slowly than the MB distribution, and are commonly modeled using a variety of empirical forms, such as the kappa distribution \cite{Livax}, the Cairns distribution \cite{Cairns}, or the $q$-Gaussian distribution \cite{Tsallisbook}. In laboratory plasmas as well, similar non-MB distributions are frequently observed and are now even viewed as exploitable features, for instance to enhance reactivity rates in thermonuclear fusion \cite{reac1,reac2,reac3}.

Although the use of empirical distributions proves effective for all practical purposes, it remains unsatisfactory for two reasons. First, it says little about the underlying dynamics and the microscopic mechanisms responsible for the emergence of these distributions. Second, the use of a single generalized non-MB distribution appears too limited to fully capture the rich diversity of observed behaviors. In this context, many different approaches have been proposed, each bringing its own perspective, ranging from generalized entropy formalisms \cite{Tsallis,Kaniadakis,Dewar} to more physically grounded descriptions based on kinetic theory \cite{Treumann,Shizgal,Shizgal2}.

A natural way to approach the problem is to view these distributions as signatures of systems that are not fully in equilibrium. Indeed, as is often the case in space plasmas, collisional processes are too rare to drive the system toward complete equilibrium within a reasonable timescale, leaving it in a quasiequilibrium state for a phenomenologically relevant period \cite{Ewar}. In such a regime, the system exhibits local equilibrium at small scales while failing to reach global equilibrium. One is then naturally led to decompose the dynamics into two scales: at small scales, the machinery of equilibrium statistical mechanics applies, whereas at larger scales the temperature is no longer uniform and develops its own dynamics. The resulting distributions can thus be understood as a superposition of statistics, in other words, \textit{superstatistics} \cite{super0}.

Such an approach is increasingly seen as a standard paradigm for understanding the emergence of non-MB distributions. It has been successfully applied to a wide range of physical systems \cite{Tur1,stur2,Tur2,stur3,Rouse,Ourabah2017,Ising,OurabahPRD,OurabahPRE,sup4,BT}, and extends naturally to complex systems beyond the realm of physics, where fluctuations and non-Gaussian behavior play a central role \cite{sup1,ss22,traffic,Market0,nature,nature2,ss4}. In plasma physics in particular, its use is becoming increasingly well established: its connections with empirical distributions known in the plasma physics community have been studied \cite{Plasma1}, it has been shown to reproduce observed features of both space and laboratory plasmas \cite{Z4,splasma3}, its implications for measurable thermodynamic quantities \cite{OurRelativistic} as well as for linear and nonlinear collective processes have been explored \cite{sup3,Omar}, and several formal results supporting its relevance have been derived \cite{Plasma2,Plasma4}. What remains largely unexplored, however, is what this approach can tell us about transport properties in nonequilibrium plasmas.

Plasmas are indeed characterized by driving gradients and forces that lead to deviations from quasi-stationary states. In sufficiently homogeneous plasmas, a macroscopic description can be adopted, in which fluxes (such as heat flux or electric currents) are related to their driving forces (such as temperature or density gradients, or electromagnetic fields) through linear relations. The corresponding proportionality coefficients are known as \textit{transport coefficients}. They are very sensitive to the underlying distribution function, and their accurate determination is crucial for predicting the evolution and stability of plasmas \cite{Braginskii1965,Balescu1988,Dum1990}. Recent studies have investigated transport coefficients, primarily using Maxwellian distributions \cite{Hagelaar2005,Lv} and various forms of the kappa distribution \cite{Du2013,Wang2017,Guo2019,Husidic2021,Husidic2022}. Here, we go a step further and provide a systematic analysis of transport coefficients in the more general setting of a quasiequilibrium plasma, where only local equilibrium holds. Although our discussion is mainly centered on the main universality classes of superstatistics, our results can be readily extended to other cases and remain valid as long as local equilibrium holds and a sufficient separation of scales allows the system’s dynamics to be decomposed into two scales.

The rest of the paper is organized as follows. In Section~\ref{SecII}, we introduce the general setting and the universality classes of superstatistics, which we compare with observed electron distributions in the solar wind to assess their validity. In Section~\ref{SecKinetic}, we derive the generic linear response of a quasiequilibrium plasma. In Section~\ref{SecIII}, we use this linear response to compute transport coefficients, such as electric and thermal conductivities, diffusion, and mobility, for the main classes of superstatistics, and discuss how they are affected by nonequilibrium conditions. In Section~\ref{SecIV}, we extend the analysis to viscous plasmas and examine how nonequilibrium conditions affect shear and bulk viscosities. Finally, in Section~\ref{SecV}, we summarize our findings and discuss possible directions for future research.

\section{Quasiequilibrium states and Superstatistics}\label{SecII}

We consider the following physical scenario: a weakly collisional plasma, as is typically the case in environments such as the solar wind. At early times, collisional effects can be neglected, and the evolution is governed primarily by the free energy stored in the system rather than by dissipative processes. This free energy can destabilize certain modes, leading to their exponential amplification from initial fluctuations. As long as the perturbation amplitudes remain small, the system is well described by \textit{linear} theory. The unstable modes, however, continue to grow until their amplitudes become large enough for nonlinear effects to set in. At this point, the linear approximation breaks down, and the system enters a \textit{nonlinear} regime, where new structures may develop.

Assuming collisions remain negligible during this stage, the system evolves under the influence of a range of nonlinear mechanisms (e.g., mode coupling, wave--particle interactions, nonlinear resonances, etc.) \cite{Treumann}. As the amplitudes of the dominant modes increase, they excite secondary (sideband) waves and introduce spatial inhomogeneities. As a consequence, the plasma evolves into a complex state characterized by a slow increase in coarse-grained entropy and by the coexistence of multiple, interdependent scales that continuously influence one another. If the nonlinear timescale remains much shorter than the collisional timescale (i.e., $\tau_{\mathrm{nl}} \ll \tau_c$), such a quasiequilibrium state may persist over a dynamically relevant time interval. In this regime, the statistical properties of the system, if probed experimentally, differ from those predicted by equilibrium statistical mechanics.

\begin{figure}[h!]
 \centering
\includegraphics[width=0.4\textwidth]{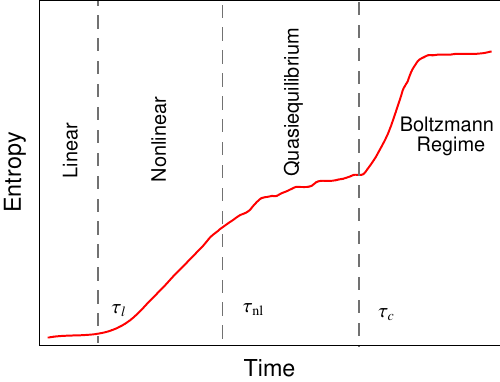}
\caption{The system undergoes four successive stages: an initial linear regime, a nonlinear phase (often referred to as violent relaxation), a long-lived quasiequilibrium state, and finally a collisional thermal equilibrium. The quasiequilibrium phase is characterized by a slow increase in coarse-grained entropy and can be effectively described within superstatistics. The final state, corresponding to maximal entropy, is governed by standard MB statistics.}
\label{Fig1}
 \end{figure}

Finally, as the evolution approaches the collisional timescale ($t \sim \tau_c$), binary collisions become increasingly significant and contribute to the irreversible relaxation of the system toward its final equilibrium state, typically described by MB statistics. This evolutionary process is qualitatively illustrated in Fig.~\ref{Fig1}.


In the quasiequilibrium regime, the system can be partitioned into small regions (cells) within which the inverse temperature $\beta$ is approximately constant\footnote{Throughout, we set $k_B=1$, so that temperature is measured in energy units and $\beta=1/T$.}. These cells represent locally equilibrated regions from which thermalization gradually spreads throughout the system. At the level of each cell, equilibrium statistical mechanics applies; in particular, the local energy distribution follows $\mathcal{P}(E|\beta) \propto \exp(-\beta E)$. However, from one cell to another, the temperature is not uniform but fluctuates. If these fluctuations occur on sufficiently long timescales, they can be described by a distribution, say $f(\beta)$. The statistical properties of the whole system are then given by a superposition of such local equilibrium statistics, namely
\begin{equation}\label{Eq1}
\mathcal{P}(E) = \int_0^{\infty} f(\beta)\, \mathcal{P}(E|\beta)\, d\beta.
\end{equation}
In equilibrium, the temperature is uniform, so that $f(\beta)=\delta(\beta-\beta_0)$, and Eq.~(\ref{Eq1}) reduces to the standard Boltzmann distribution. In general, however, the temperature varies across the system, and the form of $f(\beta)$ directly determines the distribution that is effectively measured. Although $f(\beta)$ could in principle be arbitrary, empirical evidence from a wide range of nonequilibrium systems shows that only a limited number of forms occur repeatedly. These define the three main universality classes of superstatistics:

\begin{itemize}

\item[(a)] \textbf{$\chi^2$ superstatistics:}  
In this case, the inverse temperature $\beta$ is assumed to follow a $\chi^2$ distribution with $n$ degrees of freedom,
\begin{equation}\label{f1}
f_1(\beta) = \frac{1}{\Gamma\left(\frac{n}{2}\right)} 
\left(\frac{n}{2\beta_0}\right)^{n/2} 
\beta^{n/2-1} 
\exp\left(-\frac{n\beta}{2\beta_0}\right),
\end{equation}
where $\beta_0 \equiv \langle \beta \rangle$. Starting from the local MB distribution and using Eq.~(\ref{Eq1}), the corresponding velocity distribution (in $d$ dimensions) reads
 \begin{equation}\label{B1}
    \begin{aligned}
        \mathcal{P}(v) &= \int_{0}^{\infty} d\beta f(\beta) \left(\frac{m}{2\pi \beta}\right)^{d/2} \exp\left(-\frac{\beta mv^2}{2}\right)\\
   & = \frac{\beta_0 m}{\pi^{d/2} \Gamma\left(\frac{n}{2}\right)} \frac{\Gamma\left(\frac{n+d}{2}\right)}{(1 + \frac{\beta_0 m v^2}{n})^{(n+d)/2}}. 
    \end{aligned}
    \end{equation}
This distribution exhibits power-law decay at large velocities.

\item[(b)] \textbf{Inverse-$\chi^2$ superstatistics:}  
In this case, it is the temperature $\beta^{-1}$ that follows a $\chi^2$ distribution, leading to an inverse-$\chi^2$ distribution for $\beta$,
\begin{equation}\label{f2}
f_2(\beta) = \frac{\beta_0}{\Gamma\left(\frac{n}{2}\right)} 
\left(\frac{n\beta_0}{2}\right)^{n/2} 
\beta^{-(n/2+2)} 
\exp\left(-\frac{n\beta_0}{2\beta}\right).
\end{equation}
The corresponding velocity distribution is given by
   \begin{equation}\label{B2}
    \begin{aligned}
        \mathcal{P}(v) &= \frac{2\beta_0}{\Gamma\left(\frac{n}{2}\right)} \left(\frac{m}{2\pi}\right)^{d/2} \left(\frac{n}{2\beta_0}\right)^{n/2} \left(\frac{mv^2}{\beta_0 n}\right)^{(n-d)/4}\\
        &\times \mathcal{K}_{\frac{2-d+n}{2}}\left(\sqrt{nm\beta_0}|v|\right), 
    \end{aligned}
    \end{equation}
where $\mathcal{K}_\nu$ denotes the modified Bessel function of the second kind. In contrast to the previous case, this distribution exhibits exponential decay in the high-energy tails.

\item[(c)] \textbf{Log-normal superstatistics:}  
In this case, the inverse temperature $\beta$ is assumed to follow a log-normal distribution,
\begin{equation}\label{f3}
f_3(\beta) = \frac{1}{\sqrt{2\pi}\, s\, \beta}
\exp\left(-\frac{(\ln (\beta / \mu))^2}{2s^2}\right),
\end{equation}
with mean value $\beta_0 = \mu e^{s^2/2}$. The resulting velocity distribution does not admit a closed analytical form in this case and must be evaluated numerically.

\end{itemize}

These three universality classes are not merely convenient mathematical choices; they reflect three distinct statistical scenarios that may arise at the microscopic level \cite{CLT}. The first two classes are associated with \textit{additive} random processes, acting respectively on $\beta$ or on the temperature $T \equiv 1/\beta$, whereas the third class is associated with \textit{multiplicative} random processes, typically encountered in cascading dynamics such as turbulence. Beside, their asymptotic behaviors cover the main tendencies observed in real-world data. In this sense, they provide a natural set of reference distributions, since many empirical distributions are expected to fall within one of these classes, or within simple combinations of them.

They are also supported by strong empirical evidence. In fact, the power-law decay of the first class makes it especially well suited for modeling systems with suprathermal tails, such as those commonly observed in plasmas \cite{splasma3,plasma1} and gravitational systems \cite{OurabahPRD,OurabahPRE,Mahsa}, as well as in several other physical contexts \cite{cold1,cold2,HE1,HE2}. The exponential-type decay produced by the second class has been observed in a variety of nonequilibrium systems, including  fusion plasmas \cite{Z4}, vortex glasses and liquids \cite{VG}, and harmonic oscillators interacting with a solvent bath \cite{invchi11}. Finally, the log-normal class is naturally supported by empirical data from systems where cascading processes dominate, such as turbulence \cite{stur2,stur3}. Evidence for this class has also been found in space plasma environments \cite{splasma3}, gravitational systems \cite{OurabahPRE}, and a variety of complex systems beyond physics \cite{nature,Xu,Jizba2}.

\begin{figure*}[t]
\centering

\begin{minipage}{0.3\textwidth}
\centering
\includegraphics[width=\linewidth]{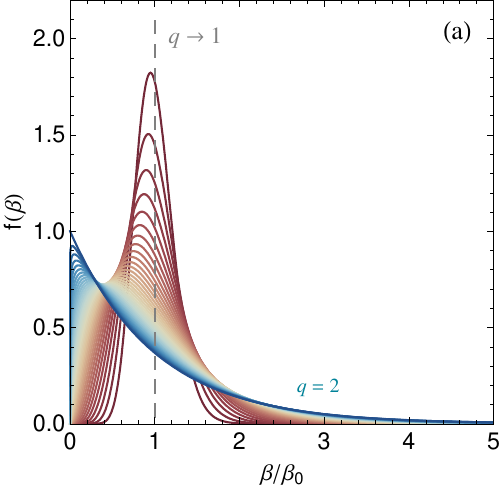}
\end{minipage}
\hfill
\begin{minipage}{0.3\textwidth}
\centering
\includegraphics[width=\linewidth]{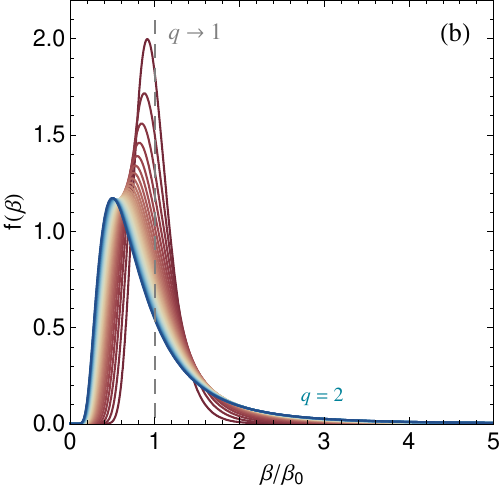}
\end{minipage}
\hfill
\begin{minipage}{0.3\textwidth}
\centering
\includegraphics[width=\linewidth]{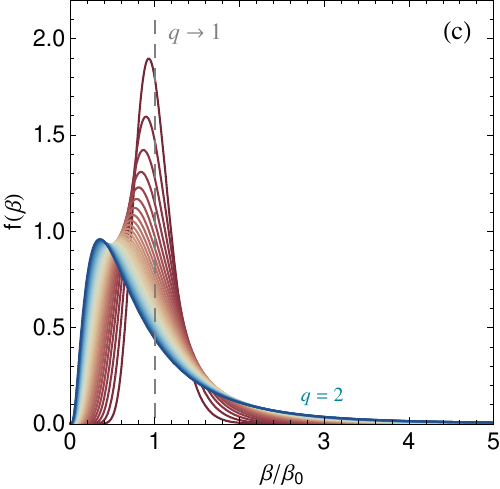}
\end{minipage}
\hfill
\begin{minipage}{0.3\textwidth}
\centering
\includegraphics[width=\linewidth]{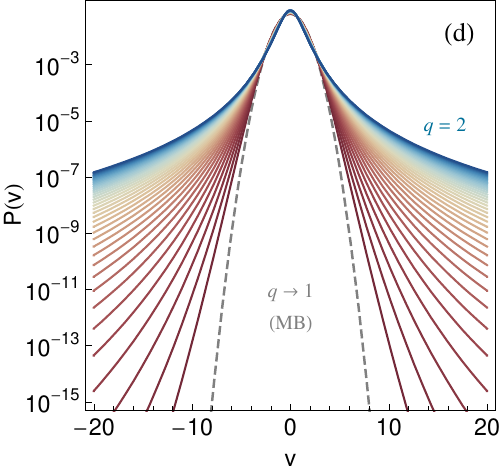}
\end{minipage}
\hfill
\begin{minipage}{0.3\textwidth}
\centering
\includegraphics[width=\linewidth]{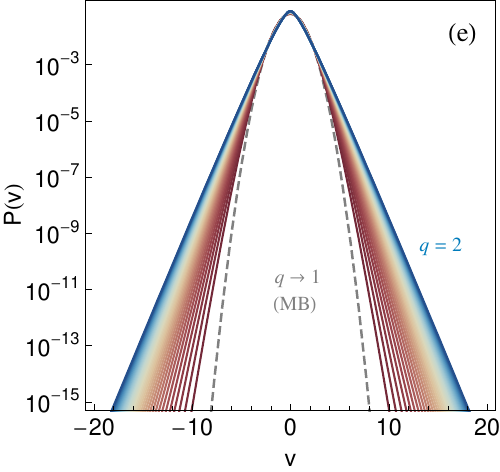}
\end{minipage}
\hfill
\begin{minipage}{0.3\textwidth}
\centering
\includegraphics[width=\linewidth]{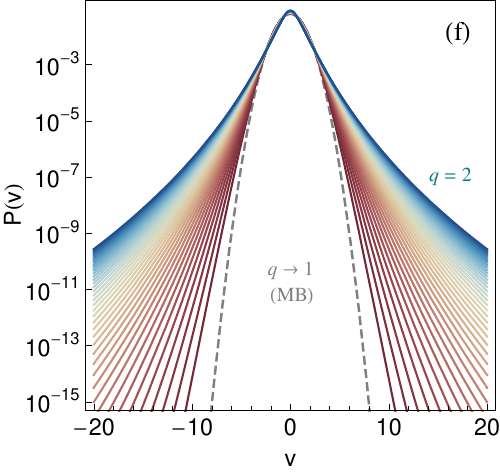}
\end{minipage}
\caption{The top panels show the three universality classes $f(\beta)$: (a) $\chi^2$, (b) inverse $\chi^2$, and (c) log-normal. The bottom panels display the corresponding velocity distributions. All distributions are parametrized by $q := \langle \beta^2 \rangle / \beta_0^2$. In the limit $q \to 1$, $f(\beta)$ collapses to a Dirac delta function, and the associated velocity distribution reduces to the Maxwellian distribution.}
\label{Fig2}
\end{figure*}

For completeness, Fig.~\ref{Fig2} displays the three distributions $f_i(\beta)$ $(i=1,2,3)$, together with their associated one-dimensional velocity distributions. To compare the three classes on the same footing, we parametrize the strength of temperature fluctuations through the dimensionless parameter
\begin{equation}
q := \frac{\langle \beta^2 \rangle}{\beta_0^2},
\end{equation}
which satisfies $q \geq 1$. For the three classes considered here, this parameter is related to the parameters of $f_i(\beta)$ as
\begin{equation}\label{qclasses}
\begin{aligned}
q_1 &= 1 + \frac{2}{n}, \\
q_2 &= \frac{n}{n-2}, \\
q_3 &= e^{s^2},
\end{aligned}
\end{equation}
corresponding respectively to the $\chi^2$, inverse-$\chi^2$, and log-normal classes. This common parametrization will be used throughout the following, since it allows the three classes to be compared in terms of the same fluctuation strength. Larger values of $q$ correspond to stronger temperature fluctuations, while the limit $q \to 1$ corresponds to the absence of fluctuations. In this limit, $f(\beta)$ reduces to $\delta(\beta-\beta_0)$ and the resulting distribution recovers the MB form.

\begin{figure*}[t]
\centering

\begin{minipage}{0.45\textwidth}
\centering
\includegraphics[width=\linewidth]{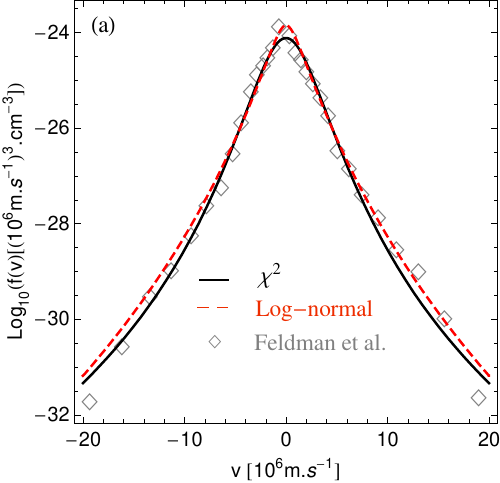}
\end{minipage}
\hfill
\begin{minipage}{0.45\textwidth}
\centering
\includegraphics[width=\linewidth]{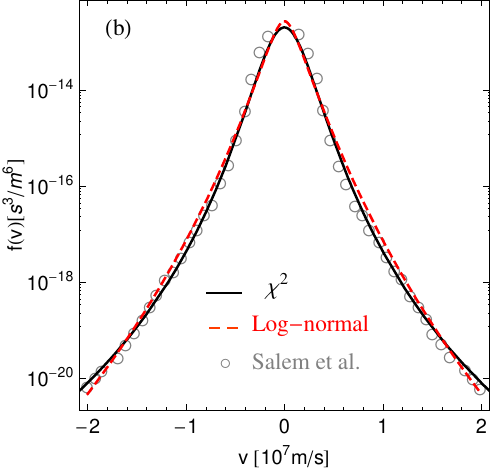}
\end{minipage}

\caption{Electron velocity distribution functions in the solar wind: (a) data from Feldman \textit{et al.} \cite{Feldman} and (b) data from Salem \textit{et al.} \cite{Salem}, together with the best fits obtained from the $\chi^2$ class (solid black line) and the log-normal class (dashed red line). The corresponding fitted parameters are given in the text.}
\label{Fig3}
\end{figure*}

As an illustration of the relevance of these distributions to weakly collisional plasmas, we compare them with observed electron velocity distributions in the solar wind. Figure~\ref{Fig3} shows the best fits obtained for two datasets, namely those of Feldman \textit{et al.} \cite{Feldman} in panel (a) and Salem et \textit{et al.} \cite{Salem} in panel (b). Only the $\chi^2$ and log-normal classes are shown. The inverse-$\chi^2$ class is omitted, since its exponential-type decay does not reproduce the observed high-energy tails.

For the Feldman \textit{et al.} dataset, shown in panel (a), the best fit obtained with the $\chi^2$ class corresponds to $q=1.28$ and $T_0 \equiv 1/\beta_0 = 1.4 \times 10^5~\mathrm{K}$, while the best fit obtained with the log-normal class corresponds to $q=2$ and $T_0 = 1.04 \times 10^5~\mathrm{K}$. In both cases, the density is fixed at $n = 33.9~\mathrm{cm}^{-3}$. For the Salem \textit{et al.} dataset, shown in panel (b), the best-fit parameters are $q=1.3$ and $T_0 = 1.2 \times 10^5~\mathrm{K}$ for the $\chi^2$ class, and $q=1.8$ and $T_0 = 1.07 \times 10^5~\mathrm{K}$ for the log-normal class, with $n = 7~\mathrm{cm}^{-3}$. In all cases, the reduced chi-squared goodness-of-fit values are $\chi^2_{\mathrm{red}} \sim 0.07$, indicating excellent agreement with the data, with deviations smaller than the estimated uncertainties. These results show that both the $\chi^2$ and log-normal classes provide an equally satisfactory description of the observed distributions.

The transport coefficients studied in the following sections depend on velocity moments of the underlying distribution. In the present superstatistical setting, these moments have a simple structure: they are obtained by averaging the MB moments over the fluctuating inverse temperature. The result is a Maxwellian-like expression multiplied by a class-dependent factor $\xi_i(\ell,q)$, which carries the entire effect of the temperature fluctuations. The detailed derivation was given in Ref.~\cite{OurRelativistic}; here we only recall the final expressions needed below. One has
\begin{equation}
\left\langle v^\ell \right\rangle_i
\equiv 
\int v^\ell \mathcal{P}_i(v)\, d^d v
=
\left\langle
\left\langle v^\ell \right\rangle_{\mathrm{MB}}
\right\rangle_{f_i(\beta)} ,
\end{equation}
where the outer average is performed over the distribution $f_i(\beta)$. Using the moments of the MB distribution together with those of $f_i(\beta)$ $(i=1,2,3)$, one obtains
\begin{equation}\label{vlvl}
\langle v^\ell \rangle_i =
\xi_i(\ell,q)
\frac{\Gamma\left( \frac{d+\ell}{2} \right)}
{\Gamma\left( \frac{d}{2} \right)}
\left( \frac{2}{\beta_0 m} \right)^{\ell/2},
\end{equation}
where the auxiliary functions are defined as
\begin{equation}\label{xii}
\begin{aligned}
\xi_1(\ell,q) & =
\frac{\Gamma\left(\frac{1}{q-1}-\frac{\ell}{2}\right)}
{\Gamma\left(\frac{1}{q-1}\right)}
\left(q-1\right)^{-\ell/2}, \\[2mm]
\xi_2(\ell,q) & =
\frac{\Gamma\left(\frac{q}{q-1}+\frac{\ell}{2}+1\right)}
{\Gamma\left(\frac{q}{q-1}\right)}
\left(\frac{q}{q-1}\right)^{-1-\ell/2}, \\[2mm]
\xi_3(\ell,q) & =
q^{\frac{\ell}{4}\left(\frac{\ell}{2}+1\right)} .
\end{aligned}
\end{equation}

Note that, for the $\chi^2$ class, the power-law tail may lead to divergent moments\footnote{This is a generic feature of distributions with power-law tails, such as $q$-Gaussian distributions or different forms of the kappa distribution. For $\int v^\ell \mathcal{P}(v)\,d^d v$ to be finite, the large-velocity decay of $\mathcal{P}(v)$ must be faster than $v^{-(\ell+d)}$, which in turn restricts the admissible parameter values of the distribution.}. Requiring finiteness of the $\ell$-th moment restricts the admissible range of $q$ to

\begin{equation}
\frac{1}{q-1}>\frac{\ell}{2}.
\end{equation}

\section{Kinetic response of a quasiequilibrium plasma}
\label{SecKinetic}

To derive the fundamental transport coefficients of the plasma, we must first determine how the electron distribution function responds to small perturbations in the electric field and in the temperature and density gradients. We begin with a kinetic description based on the Boltzmann equation
\begin{equation}
\frac{\partial f}{\partial t}+\mathbf{v} \cdot \frac{\partial f}{\partial \mathbf{r}}-\frac{e}{m}\left[\mathbf{E}+\frac{1}{c}(\mathbf{v} \times \mathbf{B})\right] \cdot  \frac{\partial f}{\partial \mathbf{v}}=C(f),
\end{equation}
where $f(\mathbf{r}, \mathbf{v}, t)$ denotes the single-particle distribution function for electrons, $\mathbf{E}$ and $\mathbf{\textbf{B}}$ are the electric and magnetic fields, and $C(f)$ is the collision term. Throughout the text, $f(\mathbf{r}, \mathbf{v}, t)$ is normalized to the density, whereas $\mathcal{P}$ is normalized to unity, so that $f= n \mathcal{P}$. In the case of a non-magnetized plasma ($\mathbf{B}=0$) under steady-state conditions ($\partial_t f=0$), the equation reduces to the stationary transport equation
\begin{equation}\label{bb}
\mathbf{v} \cdot \frac{\partial f}{\partial \mathbf{r}} - \frac{e}{m} \mathbf{E} \cdot \frac{\partial f}{\partial \mathbf{v}} = C(f).
\end{equation}

To solve this equation in the presence of weak driving forces, we use a standard linearization procedure. The distribution function is decomposed into a dominant stationary part and a small perturbation,
\begin{equation}\label{fr}
f(\mathbf{r}, \mathbf{v}, t) = f_0(\mathbf{r}, \mathbf{v}) + \delta f(\mathbf{r}, \mathbf{v}, t),
\end{equation}
with $|\delta f| \ll f_0$. In general, the collision term $C(f)$ is a nonlinear integral operator, which makes the Boltzmann equation difficult to solve exactly. However, when the electrons are sufficiently hot, electron--electron collisions may be neglected, and only interactions with the stationary ionic background need to be retained. In this regime, the collision term can be approximated by the Bhatnagar--Gross--Krook (BGK) operator \cite{BGK}
\begin{equation}
C(f)=-\nu_{ei}(v)\left(f-f_0\right)=-\nu_{ei}(v) \delta f,
\end{equation}
where $\nu_{ei}(v)$ is the electron--ion collision frequency, given by \cite{HelanderSigmar2005}
\begin{equation}\label{coll}
\nu_{\mathrm{ei}}(v)=\frac{4 \pi n_{} Z e^4 \ln \Lambda}{m_r m v^3} \approx \frac{4 \pi n_{} Z e^4 \ln \Lambda}{m^2 v^3},
\end{equation}
where $Z$ is the ion charge number, $\ln \Lambda$ is the Coulomb logarithm, and $m_r \equiv m m_i/(m+m_i) \simeq m$ is the electron--ion reduced mass. Here and in what follows, $m$ and $n$ denote the electron mass and density, respectively, unless otherwise stated.

Substituting the decomposition (\ref{fr}) into Eq.~(\ref{bb}) and neglecting second-order spatial and velocity gradients of the perturbation, we obtain
\begin{equation}\label{azerty}
\mathbf{v} \cdot \frac{\partial f_0}{\partial \mathbf{r}} - \frac{e}{m} \mathbf{E} \cdot \frac{\partial f_0}{\partial \mathbf{v}} = -\nu_{ei}(v)  \delta f.
\end{equation}
We now assume that the plasma is initially in a quasiequilibrium state, so that $f_0$ is described by a superstatistical distribution of the generic form given in Eq.~(\ref{Eq1}). This distribution reflects local temperature fluctuations around the mean value $T_0 \equiv \beta_0^{-1}$. Under this assumption, Eq.~(\ref{azerty}) yields
\begin{equation}\label{df}
\begin{aligned}
& \delta f = -   \frac{\langle \beta f_{\text{MB}}(v, \beta) \rangle_{f(\beta,\beta_0)}}{\nu_{ei}(v)} e \mathbf{E} \cdot \mathbf{v} \\&+ \frac{\langle f_{\text{MB}} (v, \beta)\rangle_{\tilde{f}(\beta, \beta_0)} }{\nu_{ei}(v) T_0^2}  \nabla T_0 \cdot \mathbf{v} 
 - \frac{ \langle f_{\text{MB}} (v, \beta)\rangle_{f(\beta,\beta_0)}}{\nu_{ei}(v) n} \mathbf{\nabla} n \cdot \mathbf{v},   
\end{aligned}
\end{equation}
where the angular brackets denote averaging with respect to the indicated weight, i.e., $\langle \bullet \rangle_{g(\beta,\beta_0)} \equiv  \int \bullet  \, g(\beta,\beta_0) d \beta $, and $\tilde{f}(\beta, \beta_0) \equiv \partial f(\beta, \beta_0) / \partial \beta_0 $. Equation~(\ref{df}) expresses the linear response of a quasiequilibrium plasma in a class-independent form: the choice of superstatistics enters only through the averages over $f(\beta)$. It therefore applies not only to the three universality classes considered here, but also to any distribution of inverse temperatures compatible with the superstatistical construction. This expression will be the basis for the calculation of the transport coefficients below.

It is useful to check that the standard Maxwellian response is recovered when temperature fluctuations are suppressed \cite{Husidic2021}. This corresponds to an infinitely narrow temperature distribution centered at $\beta_0$, namely $f(\beta, \beta_0) = \delta(\beta - \beta_0)$. In this limit, the weighted averages simplify to
\begin{equation}
\begin{aligned}
\left\langle f_{\text{MB}}(v,\beta) \right\rangle_{f(\beta, \beta_0)}
&\to f_{\text{MB}}(v, \beta_0),\\
\left\langle \beta f_{\text{MB}}(v, \beta) \right\rangle_{f(\beta, \beta_0)}
&\to \beta_0 f_{\text{MB}}(v, \beta_0),  \\
\left\langle f_{\text{MB}}(v,\beta) \right\rangle_{\tilde{f}(\beta, \beta_0)}
&\to \frac{\partial f_{\text{MB}}(v, \beta_0)}{\partial \beta_0} \\
=
&\left(\frac{3}{2\beta_0}-\frac{m v^2}{2}\right)
f_{\text{MB}}(v,\beta_0),
\end{aligned}
\end{equation}
and Eq. (\ref{df}) reduces to
\begin{equation}\label{dfM}
\delta f = - \frac{ f_{\text{MB}}(v, \beta_0)}{\nu_{ei}(v) T_0} \left [e \mathbf{E} - \frac{\nabla T_0}{T_0} (\epsilon - \langle \epsilon \rangle) - T_0\frac{\nabla n}{n}\right] \cdot \mathbf{v},
\end{equation}
where $\epsilon = m v^2 /2$ is the kinetic energy and $\langle \epsilon \rangle = 3 T_0/2$.

\section{Transport coefficients}
\label{SecIII}

We now use the linear response (\ref{df}) derived in the previous section to evaluate the main transport coefficients of the plasma. In the linear regime, the electric current density, heat flux, and particle flux are related to the electric field and to the gradients of temperature and density through macroscopic transport equations. For the model considered here, where electrons interact with a fixed neutralizing ionic background, electron--electron collisions are neglected, and no magnetic field is present, these relations read \cite{BoydSanderson2003}

\begin{equation}\label{jj}
\mathbf{j} = \sigma \mathbf{E} - \sigma \alpha \nabla T_0,
\end{equation}

\begin{equation}\label{q}
\mathbf{q} = \sigma \alpha T_0 \, \mathbf{E} - \lambda \nabla T_0,
\end{equation}

\begin{equation}\label{gamma}
\boldsymbol{\Gamma}=-D \nabla n-\mu n \boldsymbol{E}.
\end{equation}

Equation~(\ref{jj}) represents a generalized form of Ohm's law, relating the current density $\mathbf{j}$ to the electric field and to the temperature gradient. Equation~(\ref{q}) similarly extends Fourier's law, with $\mathbf{q}$ denoting the heat flux. Equation~(\ref{gamma}) corresponds to an extended Fick's law for the particle flux $\boldsymbol{\Gamma}$, accounting not only for diffusion driven by density gradients but also for the drift of charged particles under the action of the electric field. The coefficients $\sigma$, $\alpha$, $\lambda$, $D$, and $\mu$ denote, respectively, the electrical conductivity, the thermoelectric coefficient, the thermal conductivity, the diffusion coefficient, and the mobility.


Note that, in Eqs.~(\ref{jj})--(\ref{gamma}), we write the temperature as $T_0$ to emphasize its macroscopic character and to distinguish it from the local fluctuating temperature associated with each superstatistical cell. The transport equations are macroscopic relations: they describe fluxes measured over scales much larger than the size of the local cells. The temperature gradient that appears in these equations must therefore be understood as the gradient of the coarse-grained temperature, namely the mean temperature $T_0=1/\beta_0$. The local fluctuations of temperature are not ignored; rather, they enter through the distribution function and modify the velocity moments, and hence the transport coefficients themselves.

\subsection{Electric conductivity}
To obtain the electric conductivity $\sigma$, we set $\nabla T_0 = 0$ and $\nabla n = 0$ in the linear response (\ref{df}), since these gradients do not contribute to the conductivity. This gives
\begin{equation}\label{df0}
\delta f = - \frac{e}{\nu_{ei}(v)} \left\langle \beta f_{\text{MB}}(v, \beta) \right\rangle_{f(\beta,\beta_0)} \, \mathbf{E} \cdot \mathbf{v}.
\end{equation}

The current density is defined as
\begin{equation}\label{j}
\mathbf{j} = -e \int \mathbf{v}  f  d^3 v = - e \int \mathbf{v}  \delta f  d^3 v,
\end{equation}
where the contribution from the unperturbed distribution vanishes, $\int \mathbf{v} f_0 \, d^3 v = \mathbf{0}$, by parity. Substituting Eq.~(\ref{df0}) into Eq.~(\ref{j}) gives
\begin{equation}\label{jjj}
\mathbf{j} = \frac{ e^2}{3}  \mathbf{E} \int \left\langle \beta f_{\text{MB}} \right\rangle_{f(\beta,\beta_0)} \frac{v^2}{\nu_{ei}(v)} \, d^3v,
\end{equation}
where isotropy has been used through the identity
\begin{equation}
\int \mathbf{v}\mathbf{v}  F(v)  d^3 v = \frac{\mathcal{I}}{3} \int v^2 F(v)  d^3 v,
\end{equation}
with $\mathbf{v}\mathbf{v}$ denoting the dyadic product and $\mathcal{I}$ the unit tensor.

Using the collision frequency (\ref{coll}) and comparing Eq.~(\ref{jjj}) with Eq.~(\ref{jj}), we identify the electric conductivity as
\begin{equation}
 \sigma^{(q)} = \frac{ 4}{\pi Z e^2 \ln \Lambda} \sqrt{\frac{2}{m}} \beta_0^{-3/2} \xi_i(3,q),   
\end{equation}
where we have used Eq.~(\ref{vlvl}), which expresses the moments of the superstatistical distributions in terms of the auxiliary functions $\xi_i(\ell,q)$. Evaluating these functions for the three universality classes, the electric conductivity takes the explicit form
\begin{equation}\label{sigmaeq}
\left\{
\begin{aligned}
\sigma_1^{(q)} &= \frac{\Gamma\!\left[\frac{1}{q-1}-\frac{3}{2}\right]}{(q-1)^{3/2}\,\Gamma\!\left[\frac{1}{q-1}\right]} \,\sigma \quad (q<5/3), \\[6pt]
\sigma_2^{(q)} &= \left(\frac{q-1}{q} \right)^{5/2} \frac{\Gamma\!\left[\frac{1}{q-1} + \frac{7}{2}\right]}{\Gamma\!\left[\frac{q}{q-1}\right]} \,\sigma, \\[6pt]
\sigma_3^{(q)} &= q^{15/8} \,\sigma,
\end{aligned}
\right.
\end{equation}
where 
\begin{equation}
\sigma=\frac{  T_0 ^{3 / 2}}{ \sqrt{2} \pi^{3 / 2} m^{1 / 2} Z e^2 \ln \Lambda} 
\end{equation}
is the standard Maxwellian-based electric conductivity.

\begin{figure}[h!]
 \centering
\includegraphics[width=0.4\textwidth]{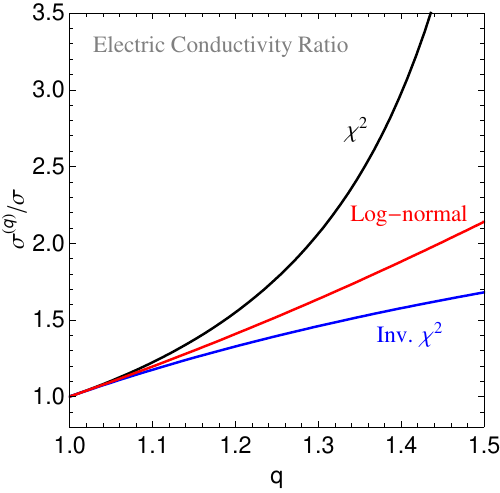}
\caption{Electric conductivity ratio $\sigma^{(q)}/\sigma$ as a function of $q := \langle \beta^2\rangle/\beta_0^2$ for the three universality classes of superstatistics.}
\label{Figsigma}
 \end{figure}

Figure~\ref{Figsigma} shows the normalized electric conductivity $\sigma^{(q)}/\sigma$ as a function of the parameter $q := \langle \beta^2 \rangle / \beta_0^2$ for the three universality classes of superstatistics. One may observe that deviations from equilibrium ($q=1$) lead to an enhancement of the conductivity for all three classes. The increase is strongest for the $\chi^2$ class, followed by the log-normal class and then the inverse-$\chi^2$ class.

\subsection{Thermoelectric coefficient}

To study the thermoelectric coefficient $\alpha$ in a superstatistical plasma, one can set $\mathbf{E}=0$ and $\nabla n=0$ in Eq.~(\ref{df}), since these terms do not affect $\alpha$. Under this condition, Eq.~(\ref{df}) reduces to
\begin{equation}\label{dft0}
\delta f = \frac{\langle f_{\text{MB}} \rangle_{\tilde{f}(\beta, \beta_0)}}{\nu_{ei}(v) T_0^2}  \nabla T_0 \cdot \mathbf{v}.
\end{equation}

The corresponding current density is then
\begin{equation}
\mathbf{j}= -\frac{e \nabla T_0}{3 T_0^2} \int \frac{\langle f_{\text{MB}} \rangle_{\tilde{f}(\beta, \beta_0)}}{\nu_{ei}(v)} v^2 \, d^3 v.
\end{equation}

Using the identity
\begin{equation} \begin{aligned} \int v^{\ell} \langle f_{\text{MB}} \rangle_{\tilde{f}(\beta, \beta_0)} d^3{v}&= \int d^3{v} v^{\ell} \int d \beta f_{\text{MB}} \frac{\partial f(\beta,\beta_0)}{\partial \beta_0} \\ &= n \frac{\partial}{\partial \beta_0} \langle v^{\ell} \rangle_i, \end{aligned} \end{equation}
the current density can be expressed explicitly as
\begin{equation}\label{j2}
\mathbf{j}= \frac{40 , \nabla T_0}{e^3 Z \ln \Lambda} \sqrt{\frac{2}{m \pi^3 \beta_0^3}} \, \xi_i(5,q).
\end{equation}

Comparing Eq.~(\ref{j2}) with Eq.~(\ref{jj}) under $\mathbf{E}=0$, and using $\sigma^{(q)}$ from Eq.~(\ref{sigmaeq}), the thermoelectric coefficient for the three universality classes reads
\begin{equation}\label{eqalphaq}
\left\{
\begin{aligned}
\alpha_1^{(q)} &= \frac{2}{7-5q} \, \alpha \quad (q<7/5), \\[2mm]
\alpha_2^{(q)} &= \frac{7q-5}{2q} \, \alpha, \\[1mm]
\alpha_3^{(q)} &= q^{5/2} \, \alpha,
\end{aligned}
\right.
\end{equation}
where
\begin{equation}
\alpha= - \frac{5}{2e}
\end{equation}
is the standard Maxwellian-based coefficient, recovered in the limit $q \to 1$.

\begin{figure}
\centering
\includegraphics[width=0.4\textwidth]{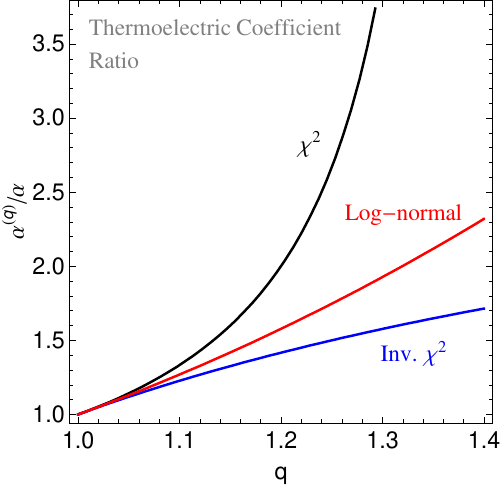}
\caption{Thermoelectric coefficient ratio $\alpha^{(q)}/\alpha$ as a function of $q$ for the three universality classes of superstatistics.}
\label{Figalpha}
\end{figure}

Figure~\ref{Figalpha} displays the normalized thermoelectric coefficient $\alpha^{(q)}/\alpha$ as a function of $q := \langle \beta^2 \rangle / \beta_0^2$ for the three universality classes. Notably, $\alpha^{(q)}$ exceeds $\alpha$ for $q>1$ across all classes, indicating that superstatistical fluctuations enhance thermoelectric coupling.

\subsection{Thermal conductivity}

To derive the thermal conductivity $\lambda$ in a superstatistical plasma, we consider the case of vanishing electric current ($\mathbf{j}=0$). In this limit, one may use the relation $\mathbf{E} = \alpha \nabla T_0$ from Eq. (\ref{jj}), which does not affect the evaluation of $\lambda$. Under this condition, the linear response (\ref{df}) reduces to

\begin{equation}\label{dfff}
\delta f = - \frac{1}{\nu_{ei}(v)} \left [e \alpha \langle \beta f_{\text{MB}} \rangle_{f(\beta,\beta_0)}- \frac{ \langle f_{\text{MB}} \rangle_{\tilde{f}(\beta, \beta_0)} }{T_0^2}\right] \nabla T_0 \cdot \mathbf{v}.
\end{equation}

Substituting Eq.~(\ref{dfff}) into Eq.~(\ref{q}) and following the same procedure as before, the heat flux can be expressed in terms of the auxiliary functions $\xi_i(\ell,q)$ as
\begin{equation}\label{qq}
\mathbf{q} = - \frac{16 \sqrt{2} T_0^{3/2}}{\pi^{3/2} Z e^4 m^{1/2} \ln \Lambda} \left [ e \alpha \xi_i(5,q)+\frac{7}{2} \xi_i(7,q)\right] \nabla T_0.
\end{equation}

The thermal conductivity then follows from Eq.~(\ref{qq}) by evaluating the auxiliary functions and using the thermoelectric coefficients (\ref{eqalphaq}) derived in the preceding subsection. For the three universality classes, it can be written as
\begin{equation} \left\{ \begin{aligned} \lambda_1^{(q)}&= \frac{2( 21 q-20)\,\Gamma\!\left[ \frac{1}{-1 + q} -\frac{7}{2}\right]} {(q-1)^{7/2}(7 q-5)\,\Gamma\!\left[\frac{1}{q-1}\right]} \lambda \quad (q<9/7), \\ \lambda_2^{(q)}&= \frac{\left(\frac{q-1}{q}\right)^{9/2}( 7 q-6)\,\Gamma\!\left[ \frac{1}{ q-1}+ \frac{9}{2}\right]} {\Gamma\!\left[\frac{1}{q-1}\right]} \lambda, \\ \lambda_3^{(q)}&= \frac{1}{2}\, q^{55/8} ( 7 q-5) \lambda. \end{aligned} \right. \end{equation}

Here,
\begin{equation}
\lambda=\frac{16 \sqrt{2}  T_0^{5 / 2}}{\pi^{3 / 2} m^{1 / 2} Z e^4 \ln \Lambda}
\end{equation}
is the standard Maxwellian thermal conductivity, recovered in the limit $q \to 1$.

\begin{figure}[h!]
\centering
\includegraphics[width=0.4\textwidth]{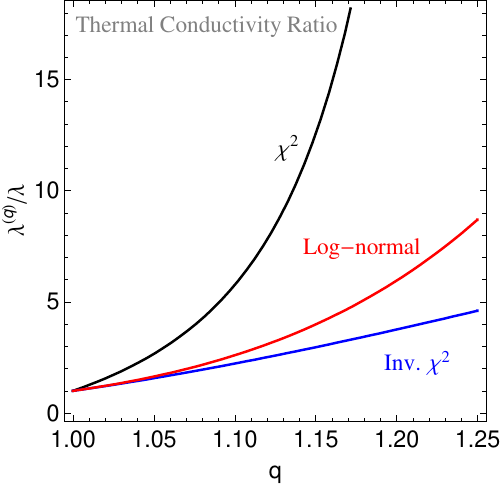}
\caption{Thermal conductivity ratio $\lambda^{(q)}/\lambda$ as a function of $q:=\langle \beta^2\rangle/\beta_0^2$, for the three universality classes of superstatistics.}
\label{Figlambda}
\end{figure}


Figure~\ref{Figlambda} shows the normalized thermal conductivity $\lambda^{(q)}/\lambda$ versus $q := \langle \beta^2 \rangle / \beta_0^2$ for the three universality classes. Fluctuations clearly increase the thermal conductivity. The enhancement of $\lambda^{(q)}$ can be attributed to the heavy tails of the distributions: energetic particles in the tails transport heat more efficiently, leading to higher thermal conductivity compared to the Maxwellian case.

\subsection{Diffusion coefficient}

The diffusion coefficient $D$ can be derived from Eq.~(\ref{df}) by setting $\mathbf{E} = 0$ and $\nabla T_0 = 0$, giving

\begin{equation}\label{dfdf}
\delta f = - \frac{f_0}{\nu_{ei}} \, \mathbf{v} \cdot \nabla n \,.
\end{equation}

The particle flux is defined as
\begin{equation}\label{Gamma1}
\mathbf{\Gamma} = \int d^3v \, \mathbf{v} f = \int d^3v \, \mathbf{v} \delta f \, ,
\end{equation}
where the integral of the unperturbed distribution vanishes, $\int d^3v \, \mathbf{v} f_0 = 0$. Substituting Eq.~(\ref{dfdf}) into Eq.~(\ref{Gamma1}) yields
\begin{equation}\label{gamma2}
\mathbf{\Gamma} = -  \int d^3v \, \mathbf{v} \mathbf{v} \frac{f_0}{\nu_{ei}} \cdot \nabla n = - \frac{1}{3 n} \langle v^2 / \nu_{ei} \rangle_i \, \nabla n.
\end{equation}

By matching the coefficients of $\nabla n$ in Eqs.~(\ref{gamma2}) and (\ref{gamma}), the diffusion coefficient can be expressed in terms of the auxiliary functions $\xi_i(\ell,q)$ as
\begin{equation}\label{DD}
D^{(q)}=\xi_i(5,q)\frac{4 \sqrt{2} \, T_0^{5 / 2}}{\pi^{3 / 2} m^{1 / 2} Z e^4 \ln \Lambda n}.
\end{equation}

Evaluating the auxiliary function explicitly gives
\begin{equation}\label{Dq} \left \{ \begin{aligned} D_1^{(q)}&= \frac{\Gamma\!\left(-\frac{5}{2} + \frac{1}{q-1}\right)}{(q-1)^{5/2}\,\Gamma\!\left(\frac{1}{q-1}\right)} D, \\ D_2^{(q)}&= \frac{(q-1)^{9/2}\,\Gamma\!\left(\frac{9}{2} + \frac{1}{q-1}\right)}{q^{7/2}\,\Gamma\!\left(\frac{1}{q-1}\right)} D, \\ D_3^{(q)}&= q^{35/8} D, \end{aligned} \right. \end{equation}
where
\begin{equation}
D=\frac{4 \sqrt{2} \, T_0^{5 / 2}}{\pi^{3 / 2} m^{1 / 2} Z e^4 \ln \Lambda n}
\end{equation}
is the standard diffusion coefficient.

\begin{figure}[h!]
\centering
\includegraphics[width=0.4\textwidth]{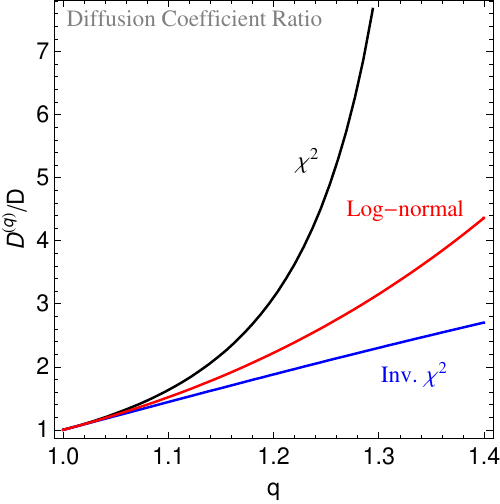}
\caption{Diffusion coefficient ratio $D^{(q)}/D$ as a function of $q:=\langle \beta^2\rangle/\beta_0^2$, for the three universality classes of superstatistics.}
\label{FigD}
\end{figure}

Figure~\ref{FigD} shows the normalized diffusion coefficient $D^{(q)}/D$ as a function of $q$ for the three universality classes. Temperature fluctuations clearly enhance particle diffusion, following the same trend observed for other transport coefficients: the $\chi^2$ class produces the strongest increase, followed by the log-normal class, and finally the inverse-$\chi^2$ class. This enhancement arises from the heavy tails of the distributions: energetic particles in the tails travel faster and contribute more effectively to the transport of matter, thereby increasing diffusion relative to the standard Maxwellian case.

\subsection{Mobility coefficient}

To derive the mobility coefficient, we set $\nabla T_0 = 0$ and $\nabla n = 0$ in Eq.~(\ref{df}), as in the derivation of the electrical conductivity $\sigma$, leading to Eq.~(\ref{df0}). Using this expression, the particle flux defined in Eq.~(\ref{Gamma1}) becomes
\begin{equation}
\mathbf{\Gamma} = - \frac{e }{3} \int d^3v \frac{v^2}{\nu_{ei}(v)} \langle f_{\text{MB}} \rangle_{f(\beta, \beta_0)} \mathbf{E}.
\end{equation}
Computing the integral and identifying with Eq.~(\ref{gamma}), we obtain the mobility as

\begin{equation}\label{muq}
\mu^{(q)}= \xi_i(3,q) \, \mu,
\end{equation}
where 
\begin{equation}
\mu=\frac{4 \sqrt{2}\, T_0^{3 / 2}}{\pi^{3 / 2} m^{1 / 2} Z e^3 \ln \Lambda n} 
\end{equation}
is the standard mobility, recovered in the limit $q \to 1$. Note that the integral involved here is identical to that appearing in the computation of the electric conductivity $\sigma$, leading to $\mu^{(q)}/\mu = \sigma^{(q)}/\sigma$. This result is expected on physical grounds, since both coefficients describe the response of the same charge carriers (electrons) to an applied electric field and thus depend on the same velocity-space moments of the distribution function. Accordingly, the standard Drude relation
\begin{equation}
\sigma^{(q)} = n e \mu^{(q)}  
\end{equation}
remains valid for all admissible values of $q$, indicating that nonequilibrium effects modify the transport coefficients while preserving the fundamental structure of charge transport.

Furthermore, combining Eqs.~(\ref{muq}) and (\ref{DD}) results in the (generalized) Einstein relation for charged particles
\begin{equation}
D^{(q)} = \frac{\xi_i(3,q)}{\xi_i(5,q)} \, \frac{ T_0 \, \mu^{(q)}}{e} \,,
\label{eq:einstein}
\end{equation}
which reduces to the standard Einstein relation for Maxwellian populations ($D=  T_0 \mu /e$) in the limit $q \to 1$.

\subsection{Application: Solar wind}

To illustrate the magnitude of superstatistical effects on the transport coefficients, we provide numerical estimates for typical solar wind conditions at 1 AU. The Maxwellian reference values are computed using standard expressions for a fully ionized plasma, with representative parameters $n = 5~\mathrm{cm^{-3}}$, $T_0 = 1.5 \times 10^5~\mathrm{K}$, and $\ln\Lambda \approx 20$. Superstatistical corrections are then incorporated through the relations derived in the previous subsections for the three universality classes. For illustration, we evaluate the coefficients at $q := \langle \beta^2 \rangle / \beta_0^2 = 1.2$. This choice is motivated by its proximity to the values inferred from solar wind data in Section~\ref{SecII}, while ensuring that all transport coefficients remain well defined (i.e., non-divergent). For consistency, we adopt the same value of $q$ for all three classes, allowing for a direct comparison of their respective impact on the transport coefficients at the same level of fluctuations. The resulting values are reported in Table~\ref{tab:transport_superstat}.

\begin{table*}[t]
\centering
\caption{Transport coefficients for electrons in the solar wind at 1 AU, comparing the Maxwellian case with the three universality classes, for $q := \langle \beta^2 \rangle / \beta_0^2 = 1.2$.}
\begin{tabular}{lcccc}
\hline\hline
Coefficient & Maxwellian & $\chi^2$ & Inverse $\chi^2$ & Log-normal \\
\hline
Electric conductivity $\sigma$ ($10^{-3}\,$ S/m) 
& $1.1$ 
& $1.7$ 
& $1.4$ 
& $1.5$ \\

Thermal conductivity $\lambda$ ($10^{-2} \,$ W m$^{-1}$ K$^{-1}$) 
& $1.8$ 
& $56.8$ 
& $6.8$ 
& $10.7$ \\

Diffusion coefficient $D$ ($10^{9} \,$ m$^2$/s) 
& $2.6$ 
& $8.1$ 
& $4.9$ 
& $5.8$ \\

Mobility $\mu$ ($10^{5} \,$ m$^2$ V$^{-1}$ s$^{-1}$) 
& $2.0$ 
& $3.1$ 
& $2.6$ 
& $2.8$ \\

Thermoelectric coefficient $\alpha$ ($10^{-8} \,$ A m$^{-1}$ K$^{-1}$) 
& $1.5$ 
& $3.0$ 
& $2.1$ 
& $2.4$ \\

\hline\hline
\end{tabular}
\label{tab:transport_superstat}
\end{table*}

The results reported in Table~\ref{tab:transport_superstat} indicate a systematic enhancement of all transport coefficients compared with their Maxwellian values. For a fixed fluctuation strength $q:= \langle \beta^2 \rangle / \beta_0^2$, the same hierarchy is observed throughout: the $\chi^2$ class produces the largest enhancement, followed by the log-normal class and then by the inverse-$\chi^2$ class. This ordering reflects the asymptotic behavior of the velocity distributions generated by the three classes: the $\chi^2$ class leads to the strongest power-law tails, the log-normal class produces an intermediate tail behavior, while the inverse-$\chi^2$ class gives a faster, essentially exponential-type decay. One also observes that, even for moderate values of $q$, the increase can be non-negligible, especially for the first class. These effects should therefore be taken into account when evaluating the transport properties of weakly collisional space plasmas, such as the solar wind, planetary magnetospheres, and the heliosheath, where high-energy overpopulations are ubiquitous and directly probed by \textit{in situ} measurements \cite{Livax}.

\section{Viscosity Coefficients}\label{SecIV}

We now extend the analysis to momentum transport and derive the corresponding viscosity coefficients. In plasmas, viscosity is particularly important because it governs the redistribution of momentum induced by velocity gradients. Through this mechanism, it contributes to the damping of waves, the dissipation of flow energy, and the stability of large-scale plasma motions in both laboratory and astrophysical settings.

Physically, viscosity arises from the transfer of momentum between neighboring fluid layers moving at different velocities. When the mean velocity of a plasma species is spatially inhomogeneous, momentum is transported from faster to slower regions, producing dissipative stresses. The derivation therefore starts from the momentum balance equation for species $\alpha$, with $\alpha=e,i$ denoting electrons and ions\footnote{In practice, the ion contribution generally dominates the viscosity, since ions carry most of the momentum.},
\begin{equation}
\frac{\partial (\rho_\alpha \mathbf{u}_\alpha)}{\partial t}
= - \nabla \cdot \left( \rho_{\alpha} \mathbf{u}_{\alpha} \mathbf{u}_{\alpha} + \mathbf{P}_{\alpha} \right)
+ \rho_{\alpha} \mathbf{f}_{\alpha}.
\end{equation}
In what follows, we omit the species index $\alpha$ to lighten the notation, with the understanding that all quantities refer to the species under consideration. Here, $\rho \mathbf{u}$ denotes the momentum density, while $\rho \mathbf{u}\mathbf{u}+\mathbf{P}$ is the corresponding momentum flux tensor, including both convective transport and internal stresses. The term $\rho \mathbf{f}$ represents the external force density acting on that species, and $\mathbf{u}=(u_x,u_y,u_z)$ denotes its mean velocity.

The central object for viscosity is the stress tensor $\mathbf{P}$. For an isotropic viscous fluid, its components can be written as \cite{Wang2018}
\begin{equation}\label{Pij}
P_{ij} = p\,\delta_{ij}
- \eta \left( \frac{\partial u_i}{\partial r_j} + \frac{\partial u_j}{\partial r_i}
- \frac{2}{3} \delta_{ij} \nabla \cdot \mathbf{u} \right)
- \zeta \, \delta_{ij} \, \nabla \cdot \mathbf{u},
\end{equation}
where $p$ is the static pressure and $\delta_{ij}$ is the Kronecker delta. This decomposition separates the isotropic pressure contribution from the dissipative stresses induced by velocity gradients. The coefficient $\eta$ measures the response to shear deformations, while $\zeta$ measures the response to compression or expansion, through $\nabla\cdot\mathbf{u}$. In Einstein summation convention,
\begin{equation}
\nabla \cdot \mathbf{u} = \frac{\partial u_k}{\partial r_k},
\end{equation}
where a repeated index $k=x,y,z$ implies summation over the three spatial directions.

At the kinetic level, the same stress tensor is obtained from the second central moment of the distribution function,
\begin{equation}\label{Pijc}
P_{ij} = m \int  (v_i - u_i)(v_j - u_j)\, f(\mathbf{r},\mathbf{v},t)\, d^3{v}.
\end{equation}
The viscosity coefficients are then extracted by evaluating Eq.~(\ref{Pijc}) for superstatistical distributions and comparing the resulting tensor with the hydrodynamic form (\ref{Pij}). This comparison directly reveals how nonequilibrium effects modify the shear and bulk viscosities.

\subsection{Shear viscosity}

To extract the shear viscosity, we focus on an off-diagonal component of the stress tensor, say $P_{xz}$. From Eq.~(\ref{Pij}), one has
\begin{equation}
P_{xz} = - \eta \left( \frac{\partial u_x}{\partial z} + \frac{\partial u_z}{\partial x} \right),
\end{equation}
where $\eta$ is the shear viscosity coefficient. At the kinetic level, the same component is expressed in terms of the distribution function as
\begin{equation}\label{Pxzc}
P_{xz} = m \int  \left( v_x - u_x \right)\left( v_z - u_z \right)
\, f(\mathbf{r},\mathbf{v},t)\, d^3{v}.
\end{equation}

Using the time-independent kinetic equation (\ref{azerty}), the distribution function can be written as
\begin{equation}\label{ff}
f \equiv f_0+ \delta f= f_0
- \frac{\mathbf{v}}{\nu} \cdot \frac{\partial f_{0}}{\partial \mathbf{r}} 
- \frac{Q \mathbf{E}}{m \nu} \cdot \frac{\partial f_0{}}{\partial \mathbf{v}} \,,
\end{equation}
where $Q$ is the charge of the species under consideration, with $Q=-e$ for electrons and $Q=+Ze$ for ions, and $\nu$ is an effective relaxation frequency. Accordingly, the $xz$-component of the stress tensor becomes
\begin{equation}\label{PPxz}
P_{xz}
= m \int d^3{v} \, (v_x - u_x)(v_z - u_z)
\left[
f_{0}
- \frac{\mathbf{v} }{\nu} \cdot \frac{\partial f_{0}}{\partial \mathbf{r}}
+ \frac{Q \mathbf{E}}{m \nu} \cdot \frac{\partial f_{0}}{\partial \mathbf{v}}
\right].
\end{equation}

For a distribution isotropic around the mean velocity $\mathbf{u}$, the first term in the integrand of Eq.~(\ref{PPxz}) does not contribute, by parity. The field-driven term also vanishes after integration for the same symmetry reason. The only contribution to $P_{xz}$ therefore comes from the spatial-gradient term, and Eq.~(\ref{PPxz}) reduces to
\begin{equation}\label{PPPxz}
P_{xz}
= - \frac{m}{\nu} \int d^3{v} \,
(v_x - u_x)(v_z - u_z)\,
\mathbf{v} \cdot \frac{\partial f_{0}}{\partial \mathbf{r}}.
\end{equation}

For a superstatistical distribution $f_0$ of the form given in Eq.~(\ref{Eq1}), the order of integration can be exchanged, giving
\begin{equation}\label{aaa1}
\begin{aligned}
& P_{xz}
= - \frac{m}{\nu} \left \langle \int d^3{v} \,
(v_x - u_x)(v_z - u_z)\,
\mathbf{v} \cdot \frac{\partial f_{\text{MB}}}{\partial \mathbf{r}} \right \rangle_{f(\beta,\beta_0)} \\
&=  - \frac{m^2}{\nu} \left \langle \beta \int d^3{v} \,
(v_x - u_x)(v_z - u_z)\,
\mathcal{A}(\mathbf{r},\mathbf{v}) f_{\text{MB}} \right \rangle_{f(\beta,\beta_0)},
\end{aligned}
\end{equation}
where
\begin{equation}\label{Arv}
\mathcal{A}(\mathbf{r}, \mathbf{v})
= \sum_{k = x,y,z} (v_k - u_k)\,
\mathbf{v} \cdot \frac{\partial u_k}{\partial \mathbf{r}}.
\end{equation}

After evaluating the integral, as detailed in Appendix~\ref{AppA}, the $xz$-component of the stress tensor becomes
\begin{equation}\label{aap}
P_{xz}= - \frac{ \xi_i(2,q) n  T_0 }{\nu}
\left(
\frac{\partial u_x}{\partial z}
+ \frac{\partial u_z}{\partial x}
\right).
\end{equation}
Comparison with Eq.~(\ref{Pij}) then gives the shear viscosity. Evaluating the auxiliary function $\xi_i(2,q)$ for the three universality classes, one obtains
\begin{equation}\label{etaq} \left \{ \begin{aligned} \eta_1^{(q)}&= \frac{1}{2-q} \eta \quad (q<2), \\ \eta_2^{(q)}&= \frac{2q-1}{q}\eta, \\ \eta_3^{(q)}&= q \eta, \end{aligned} \right. \end{equation}
where
\begin{equation}
\eta = \frac{n T_0}{\nu}
\end{equation}
is the shear viscosity coefficient for a Maxwellian distribution, recovered in the limit $q \to 1$.

\subsection{Bulk viscosity}

We now turn to the bulk viscosity, which is associated with the isotropic part of the stress tensor and therefore with the diagonal components. While the shear viscosity was extracted above from the off-diagonal component $P_{xz}$, the bulk viscosity is obtained from the trace of the stress tensor. In plasma hydrodynamics, the average of the three diagonal components is written as
\begin{equation}\label{paz}
\frac{1}{3}\left({P}_{xx} + {P}_{yy} + {P}_{zz}\right) = p - \zeta \, \nabla \cdot \mathbf{u},
\end{equation}
where $p$ is the static thermodynamic pressure and $\zeta$ is the bulk viscosity coefficient.

\begin{figure}[h!]
\centering
\includegraphics[width=0.4\textwidth]{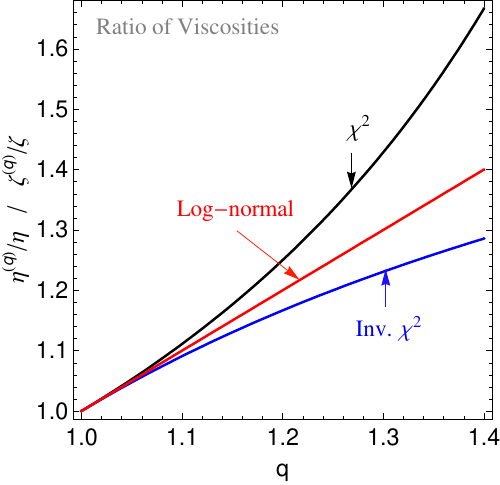}
\caption{Shear- and bulk-viscosity ratio, $\eta^{(q)}/\eta = \zeta^{(q)}/\zeta$, as a function of $q := \langle \beta^2\rangle / \beta_0^2$ for the three universality classes of superstatistics.}
\label{FigV}
\end{figure}

Focusing on a single diagonal component, say $P_{xx}$, it can be expressed in terms of the particle distribution as
\begin{equation}
P_{xx} = \int d^3 v \, m (v_x - u_x)^2 f(\mathbf{r},\mathbf{v}).
\label{Pxx_int}
\end{equation}
Substituting the explicit form of $f(\mathbf{r},\mathbf{v})$ from Eq.~(\ref{ff}), this becomes
\begin{equation}\label{Pxx}
\begin{aligned}
P_{xx} &= m\int d^3{v} \,  (v_x - u_x)^2 
\left[
1
- \frac{1}{\nu}
\left(
\mathbf{v} \cdot \frac{\partial}{\partial \mathbf{r}}
+ \frac{q \mathbf{E}}{m} \cdot \frac{\partial}{\partial \mathbf{v}}
\right)
\right] f_{0} \\
&= p 
- \frac{m}{\nu} \int d^3{v} \, (v_x - u_x)^2
\left[
\mathbf{v} \cdot \frac{\partial f_{0}}{\partial \mathbf{r}} 
+ \frac{q \mathbf{E}}{m} \cdot \frac{\partial f_{0}}{\partial \mathbf{v}} 
\right] \\
&= p 
- \frac{m}{\nu} \int d^3{v} \, (v_x - u_x)^2
\mathbf{v} \cdot \frac{\partial f_{0}}{\partial \mathbf{r}} \, ,
\end{aligned}
\end{equation}
where the field-driven term does not contribute by symmetry. Under the assumption of isotropy, the static pressure is given by
\begin{equation}
p = m \int d^3{v} \, (v_x - u_x)^2 \, f_0
= \frac{m}{3} \int d^3{v} \, (\mathbf{v} - \mathbf{u})^2 \, f_0.
\end{equation}
For the three universality classes of superstatistics, this gives
\begin{equation}\label{p0}
p= \xi_i(2,q) n T_0,
\end{equation}
while the second term in the last line of Eq.~(\ref{Pxx}) becomes (see Appendix~\ref{AppB})
\begin{equation}\label{PPx}
\begin{aligned}
&- \frac{m}{\nu} \int d^3{v} \, (v_x - u_x)^2
\mathbf{v} \cdot \frac{\partial f_{0}}{\partial \mathbf{r}} \\
&= - \frac{m}{\nu} \left \langle \int d^3{v} \,
(v_x - u_x)^2\,
\mathbf{v} \cdot \frac{\partial f_{\text{MB}}}{\partial \mathbf{r}} \right \rangle_{f(\beta,\beta_0)} \\
&=  - \frac{m^2}{\nu} \left \langle \beta \int d^3{v} \,
(v_x - u_x)^2\,
\mathcal{A}(\mathbf{r},\mathbf{v}) f_{\text{MB}} \right \rangle_{f(\beta,\beta_0)}\\
&= -\frac{\xi_i(2,q)  n T_0}{\nu} \left( 3\frac{\partial u_x}{\partial x} + \frac{\partial u_y}{\partial y} + \frac{\partial u_z}{\partial z} \right).
\end{aligned}
\end{equation}
Substituting Eqs.~(\ref{PPx}) and (\ref{p0}) into Eq.~(\ref{Pxx}) yields
\begin{equation}
P_{xx} = 2 n  T_0 \xi_i(2,q) \left[ 1 - \frac{1}{\nu} \left( 3\frac{\partial u_x}{\partial x} + \frac{\partial u_y}{\partial y} + \frac{\partial u_z}{\partial z} \right) \right].
\end{equation}
Due to the symmetry of the stress tensor, the $yy$- and $zz$-components follow by cyclic permutation of the velocity-gradient terms. Identifying the trace of the resulting tensor with Eq.~(\ref{paz}), the bulk viscosity coefficient is obtained as
\begin{equation}
\zeta ^{(q)}= \xi_i(2,q) \, \frac{5 n  T_0}{3 \nu}.
\end{equation}
Thus, the bulk and shear viscosities are modified by the same superstatistical factor,
\begin{equation}
\frac{\zeta^{(q)}}{\zeta} = \frac{\eta^{(q)}}{\eta} = \xi_i(2,q).
\end{equation}
Figure~\ref{FigV} shows this common ratio as a function of $q := \langle \beta^2 \rangle / \beta_0^2$ for the three universality classes. In all cases, the ratio is larger than unity, showing that superstatistical effects enhance both shear and bulk viscosities relative to the Maxwellian case. This enhancement reflects the high-energy overpopulation induced by the non-Maxwellian tails, which increases momentum transport across neighboring fluid layers.

\section{Conclusions} \label{SecV}

In this paper, we investigated plasmas in quasiequilibrium states whose statistical properties can be described as a continuous superposition of Maxwellian distributions, namely within superstatistics. Such states are naturally associated with non-Maxwellian distributions exhibiting heavy tails, as supported by countless \textit{in situ} measurements in space plasmas and by observations in laboratory plasmas. Starting from a kinetic treatment based on the Boltzmann equation, and considering linear perturbations around a quasiequilibrium state, we derived macroscopic relations connecting fluxes to their driving forces, with superstatistical effects entering through the corresponding transport coefficients. Since these coefficients determine the transport of charge, heat, particles, and momentum, their accurate evaluation is essential for describing the macroscopic behavior of nonequilibrium plasmas.

Our results show that departures from the equilibrium Maxwellian distribution lead to a systematic enhancement of all transport coefficients. This enhancement can be interpreted as a consequence of the increased population of energetic particles in the non-Maxwellian tails, which strengthens the transport of charge, heat, particles, and momentum in the plasma. Quantitatively, comparing the three universality classes of superstatistics reveals a clear hierarchy: for the same level of temperature fluctuations, the $\chi^2$ class produces the strongest effect, followed by the log-normal class and then by the inverse-$\chi^2$ class. This ordering reflects the asymptotic behavior of the velocity distributions generated by the three classes, ranging from strong power-law tails, to intermediate tails, and finally to exponential-type decay. It therefore provides a direct link between the microscopic structure of the distribution tail and the magnitude of the macroscopic transport response.

This work naturally opens several perspectives for future research. In particular, the results derived here could be generalized to more realistic scenarios. One possible extension is to include the effect of magnetic fields, which have been neglected in the present treatment. In that case, the transport coefficients become anisotropic, with parallel, perpendicular, and Hall components for conductivity, heat flux, diffusion, and viscosity \cite{Hagelaar2007}. Another direction would be to go beyond the simplified collision operator used here and consider more realistic Coulomb collision operators, especially Landau or Fokker--Planck operators (see, e.g., Refs.~\cite{Risken1996,operator,Ourabahbook} and references therein). On the observational side, it would be important to confront the present predictions with estimates of the transport coefficients obtained from direct numerical integration of measured distribution functions. Finally, the present analysis could also be extended to situations in which the level of fluctuations, encoded in $q$, varies in space and time (see Ref.~\cite{OurabahPRE2024} for a dynamical formulation of superstatistics). This would provide a more flexible description of nonequilibrium plasmas, in which the deviation from Maxwellian equilibrium can evolve in response to the plasma dynamics.

\onecolumngrid
\appendix

\section{}\label{AppA}

In this Appendix, we derive Eq.~(\ref{aap}) by explicitly evaluating the velocity integral entering the stress tensor. Starting from Eq.~(\ref{aaa1}), the $xz$-component of the stress tensor is written as
\begin{equation}
P_{xz}=  - \frac{m^2}{\nu} \langle \beta \, \mathcal{I} \rangle_{f(\beta,\beta_0)},
\end{equation}
where
\begin{equation}
\mathcal{I} :=  \int_{\mathbb{R}^3} d^3{v} \,
(v_x - u_x)(v_z - u_z)\,
\mathcal{A}(\mathbf{r},\mathbf{v}) f_{\text{MB}}.
\end{equation}

Substituting the expression of $\mathcal{A}(\mathbf{r},\mathbf{v})$ from Eq.~(\ref{Arv}), one obtains
\begin{equation}
\begin{aligned}
\mathcal{I} &=  \int_{\mathbb{R}^3} d^3{v} \,
(v_x - u_x)(v_z - u_z)\,
\left [(v_x-u_x) v_z \frac{\partial u_x}{\partial z} + (v_z -u_z) \frac{\partial u_z}{ \partial x}\right] f_{\text{MB}}.
\end{aligned}
\end{equation}

Exploiting the isotropy of the Maxwell--Boltzmann distribution around $\mathbf{u}$, only even contributions in $(\mathbf{v}-\mathbf{u})$ survive upon integration, yielding
\begin{equation}
\begin{aligned}
\mathcal{I}
&=    \left(\frac{\partial u_x}{\partial z} + \frac{\partial u_z}{\partial x} \right)
\int_{\mathbb{R}^3} d^3v \, (v_x - u_x)^2 (v_z - u_z)^2 \,  f_{\text{MB}}.
\end{aligned}
\end{equation}

Introducing spherical coordinates in velocity space and performing the angular integrations, the above expression reduces to
\begin{equation}
\begin{aligned}
\mathcal{I}
&=   \left(\frac{\partial u_x}{\partial z} + \frac{\partial u_z}{\partial x} \right)
\underbrace{\int_{0}^{\pi} d \phi \int_0^{2 \pi}d \theta \int_0^{\infty} dv \,
v^6 \sin^3 \theta \cos^2 \theta \cos^2 \phi  f_{\text{MB}}}_{=\, n \, {\langle v^4 \rangle}/{15}}.
\end{aligned}
\end{equation}

Therefore,
\begin{equation}
P_{xz}=  - \frac{m^2}{15 \nu}  \left(\frac{\partial u_x}{\partial z} + \frac{\partial u_z}{\partial x} \right) \langle \beta v^4\rangle_{f(\beta, \beta_0)}.
\end{equation}

Finally, using the definition of the auxiliary functions, Eq.~(\ref{aap}) of the main text is readily recovered.

\section{}\label{AppB}

In this Appendix, we explicitly evaluate the integral to recover the final form of Eq.~(\ref{PPx}). The third line of Eq.~(\ref{PPx}) can be expressed as

\begin{equation}\label{apl}
P_{xx}= - \frac{m^2}{\nu} \, \langle \beta \, \mathcal{J} \rangle_{f(\beta, \beta_0)}, 
\end{equation}
where
\begin{equation}
\mathcal{J} := \int_{\mathbb{R}^3} d^3 v \,
(v_x - u_x)^2 \,
\mathcal{A}(\mathbf{r},\mathbf{v}) f_{\rm MB}.
\end{equation}

Substituting the expression of $\mathcal{A}(\mathbf{r}, \mathbf{v})$ from Eq.~(\ref{Arv}), the integral becomes
\begin{equation}
\mathcal{J} = \int_{\mathbb{R}^3} d^3 v \, (v_x - u_x)^2 \sum_{k=x,y,z} (v_k - u_k) \sum_{l=x,y,z} v_l \frac{\partial u_k}{\partial r_l} \, f_{\rm MB}.
\end{equation}

Introducing the velocity fluctuations $\mathbf{c} := \mathbf{v} - \mathbf{u}$, so that $v_l = c_l + u_l$, the integrand can be written as
\begin{equation}
\sum_{k,l} c_x^2 \, c_k \, (c_l + u_l) \, \frac{\partial u_k}{\partial r_l}.
\end{equation}

By isotropy of the Maxwell-Boltzmann distribution, only terms with even powers in each velocity component survive. This leaves only the contributions with $k = l$, yielding $c_x^2 c_k^2 \, \partial u_k / \partial r_k$. All cross terms ($k \ne l$) and terms proportional to $u_l$ vanish.  Hence, the integral reduces to
\begin{equation}
\mathcal{J} = \sum_{k=x,y,z} \frac{\partial u_k}{\partial r_k} \int_{\mathbb{R}^3} d^3 c \, c_x^2 c_k^2 f_{\rm MB}.
\end{equation}

The remaining integral is a standard Gaussian moment of the Maxwell-Boltzmann distribution:
\begin{equation}
\int_{\mathbb{R}^3} d^3 c \, c_x^4 f_{\rm MB} = \frac{3}{\beta^2 m^2}, \quad 
\int_{\mathbb{R}^3} d^3 c \, c_x^2 c_y^2 f_{\rm MB} = \int_{\mathbb{R}^3} d^3 c \, c_x^2 c_z^2 f_{\rm MB} = \frac{1}{\beta^2 m^2}.
\end{equation}

Thus, we obtain
\begin{equation}
\mathcal{J} = \left[3 \frac{\partial u_x}{\partial x} + \frac{\partial u_y}{\partial y} + \frac{\partial u_z}{\partial z}\right] \left(\frac{1}{\beta m}\right)^2.
\end{equation}

Finally, substituting in Eq.~(\ref{apl}) and using the definition of the auxiliary functions, the last line of Eq.~(\ref{PPx}) in the main text is readily recovered.

\twocolumngrid


\begin{thebibliography}{00}
\bibitem{cold1} E. Lutz, Anomalous diffusion and Tsallis statistics in an optical lattice, \href{https://doi.org/10.1103/PhysRevA.67.051402}{Phys. Rev. A. \textbf{67}, 051402(R) (2003).} 

 

\bibitem{cold2} P. Douglas, S. Bergamini, and F. Renzoni, Tunable Tsallis Distributions in Dissipative Optical Lattices, \href{https://doi.org/10.1103/PhysRevLett.96.110601}{Phys. Rev. Lett. \textbf{96}, 110601 (2006).} 
\bibitem{HE1} V. Khachatryan \textit{et al.} (CMS Collaboration), Transverse-Momentum and Pseudorapidity Distributions of Charged Hadrons in $pp$ Collisions at $\sqrt{s}=7 \operatorname{TeV}$, \href{https://doi.org/10.1103/PhysRevLett.105.022002}{Phys. Rev. Lett. \textbf{105}, 022002 (2010).} 

\bibitem{HE2} A. Adare \textit{et al.} (PHENIX Collaboration), Measurement of neutral mesons in $p+p$ collisions at $\sqrt{s}=200 \operatorname{GeV}$ and scaling properties of hadron production, \href{https://doi.org/10.1103/PhysRevD.83.052004}{Phys. Rev. D \textbf{83}, 052004 (2011).}

\bibitem{azerty1} S. Olbert, in \textit{Physics of the Magnetosphere}, Astrophysics and
Space Science Library, Vol. 10, edited by R. D. L. Carovillano and J. F. McClay (D. Reidel, Dordrecht, 1968), p. 641.


\bibitem{azerty2} V. M. Vasyliunas, A survey of low-energy electrons in the evening sector of the magnetosphere with OGO 1 and OGO 3, \href{ https://doi.org/10.1029/JA073i009p02839}{J. Geophys. Res. \textbf{73}, 2839 (1968).} 

\bibitem{azerty3} M. Maksimovic, V. Pierrard, and P. Riley, Ulysses electron distributions fitted with Kappa functions, \href{ https://doi.org/10.1029/97GL00992}{Geophys. Res. Let. \textbf{24}, 1151 (1997).} 

\bibitem{azerty4} I. Zouganelis, Measuring suprathermal electron parameters in space plasmas: Implementation of the quasi-thermal noise spectroscopy with kappa distributions using in situ Ulysses/URAP radio measurements in the solar wind, \href{ https://doi.org/10.1029/2007JA012979}{J. Geophys. Res. \textbf{113}, A08111 (2008).} 

\bibitem{azerty5} G. Livadiotis and D. J. McComas, Beyond kappa distributions: Exploiting Tsallis statistical mechanics in space plasmas, \href{ https://doi.org/10.1029/2009JA014352}{J. Geophys. Res. \textbf{114}, A11105 (2009).}





\bibitem{Astro1} J. C. Carvalho, J. D. do Nascimento, Jr., R. Silva, and J. R. De Medeiros, Non-Gaussian statistics and stellar rotational velocities of main-sequence field, 
\href{https://doi.org/10.1088/0004-637X/696/1/L48}{Astrophys. J. \textbf{696}, L48 (2009).} 



\bibitem{Astro3} B. B. Soares and J. R. P. Silva, On the rotation of ONC stars in the Tsallis formalism context, \href{https://doi.org/10.1209/0295-5075/96/19001}{EPL \textbf{96}, 19001 (2011).}







\bibitem{galaxy} A. L. B. Ribeiro, P. A. A. Lopes, and M. Trevisan, Non-Gaussian velocity distributions — the effect on virial mass estimates of galaxy groups, \href{https://doi.org/10.1111/j.1745-3933.2011.01038.x}{MNRAS \textbf{413}, L81 (2011).}

\bibitem{Astro4} M. Curé, D. F. Rial, A. Christen, and J. Cassetti, A method to deconvolve stellar rotational velocities, \href{https://doi.org/10.1051/0004-6361/201323344}{Astronom. Astrophys. \textbf{565}, A85 (2014).}



\bibitem{Livax} G. Livadiotis, \textit{Kappa Distributions: Theory and Applications in Plasmas} (Elsevier Science, New York, 2017).

\bibitem{Cairns} R. A Cairns, \textit{et al.}, Electrostatic solitary structures in non-thermal plasmas, \href{https://doi.org/10.1029/95GL02781}{Geophys. Res. Lett. \textbf{22}, 2709 (1995).}

\bibitem{Tsallisbook} C. Tsallis, \textit{Introduction to Nonextensive Statistical Mechanics: Approaching a Complex World} (Springer, New York, 2009).


\bibitem{reac1} B. I. Squarer, C. Presilla, and R. Onofrio, 
Enhancement of fusion reactivities using non-Maxwellian energy distributions, \href{https://doi.org/10.1103/PhysRevE.109.025207}{Phys. Rev. E \textbf{109}, 025207 (2024).}


\bibitem{reac2} K. Ourabah,
Reaction rates in quasiequilibrium states, \href{https://doi.org/10.1103/PhysRevE.111.034115}{Phys. Rev. E \textbf{111}, 034115 (2025).}


\bibitem{reac3} Y. Ye, W. Zhang, and B. Wan, Enhancement of the fusion reactivity due to the D-T non-Maxwellian ion distribution and its impact on Lawson criterion, \href{https://doi.org/10.1063/5.0276381}{Phys. Plasmas \textbf{32}, 092504 (2025).}


















\bibitem{Tsallis} C. Tsallis, Possible generalization of Boltzmann-Gibbs statistics, \href{https://link.springer.com/article/10.1007/bf01016429}{J. Stat. Phys. \textbf{52}, 479 (1988).}

\bibitem{Kaniadakis} G. Kaniadakis, Statistical mechanics in the context of special relativity, \href{https://doi.org/10.1103/PhysRevE.66.056125}{Phys. Rev. E \textbf{66}, 056125 (2002).}


\bibitem{Dewar} R. C. Dewar, Information theory explanation of the fluctuation theorem, maximum entropy production and self-organized criticality in non-equilibrium stationary states, \href{https://doi.org/10.1088/0305-4470/36/3/303}{J. Phys. A \textbf{36}, 631 (2003).} 



\bibitem{Treumann} R. A. Treumann, Kinetic Theoretical Foundation of Lorentzian Statistical
Mechanics, \href{https://doi.org/10.1238/Physica.Regular.059a00019}{Phys. Scr. \textbf{59}, 19 (1999).}

\bibitem{Shizgal} B. D. Shizgal, Kappa and other nonequilibrium distributions from the Fokker-Planck equation and the
relationship to Tsallis entropy, \href{https://doi.org/10.1103/PhysRevE.97.052144}{Phys. Rev. E \textbf{97}, 052144 (2018).}

\bibitem{Shizgal2} A. D. Oylukan and B. D. Shizgal, Nonequilibrium distributions from the Fokker-Planck equation: Kappa distributions and Tsallis entropy, \href{https://doi.org/10.1103/PhysRevE.108.014111}{Phys. Rev. E \textbf{108}, 014111 (2023).}




\bibitem{Ewar} R. J. Ewart \textit{et al.}, \href{https://doi.org/10.1073/pnas.2417813122}{PNAS \textbf{122}, 2417813122 (2025).}



\bibitem{super0} C. Beck and E. G. D. Cohen, Superstatistics, \href{https://doi.org/10.1016/S0378-4371(03)00019-0}{Physica A \textbf{322}, 267 (2003).}










\bibitem{Tur1} C. Beck, Statistics of Three-Dimensional Lagrangian Turbulence, \href{https://journals.aps.org/prl/abstract/10.1103/PhysRevLett.98.064502}{Phys. Rev. Lett \textbf{98}, 064502 (2007).}
\bibitem{stur2} A. Reynolds, Superstatistical Mechanics of Tracer-
Particle Motions in Turbulence, \href{https://doi.org/10.1103/PhysRevLett.91.084503}{Phys. Rev. Lett. \textbf{91},
084503 (2003).} 

\bibitem{Tur2} M. Mehrafarin, Superstatistics as the statistics of quasiequilibrium states: Application to fully developed turbulence, \href{https://doi.org/10.1103/PhysRevE.84.022102}{Phys. Rev. E \textbf{84}, 022102 (2011).}
\bibitem{stur3} S. Jung and H. L. Swinney, Velocity difference statistics
in turbulence, \href{https://doi.org/10.1103/PhysRevE.72.026304}{Phys. Rev. E \textbf{72}, 026304 (2005).} 
\bibitem{Rouse} I. Rouse and S. Willitsch, Superstatistical Energy Distributions of an Ion in an Ultracold Buffer Gas, \href{https://doi.org/10.1103/PhysRevLett.118.143401}{Phys. Rev. Lett. \textbf{118}, 143401 (2017).}

\bibitem{Ourabah2017} K. Ourabah, Quantum entanglement and temperature fluctuations, \href{https://doi.org/10.1103/PhysRevE.95.042111}{Phys. Rev. E \textbf{95}, 042111 (2017).}

\bibitem{Ising} J. Cheraghalizadeh, M. Seifi, Z. Ebadi, H. Mohammadzadeh, and M. N. Najaf, Superstatistical two-temperature Ising model, \href{https://doi.org/10.1103/PhysRevE.103.032104}{Phys. Rev. E \textbf{103}, 032104 (2021).}


\bibitem{OurabahPRD} K. Ourabah, Quasiequilibrium self-gravitating systems, \href{https://doi.org/10.1103/PhysRevD.102.043017}{Phys. Rev. D \textbf{102}, 043017 (2020).}

\bibitem{OurabahPRE} K. Ourabah, Generalized statistical mechanics of stellar systems, \href{https://doi.org/10.1103/PhysRevE.105.064108}{Phys. Rev. E \textbf{105}, 064108 (2022).}

\bibitem{sup4} G. C. Yalcin and C. Beck, Generalized statistical mechanics of cosmic rays: Application to positron-electron spectral indices, \href{https://doi.org/10.1038/s41598-018-20036-6}{Sci. Rep. \textbf{8}, 1764 (2018).}

\bibitem{BT} C. Beck and C. Tsallis, Anomalous velocity distributions in slow quantum-tunneling chemical reactions, \href{https://doi.org/10.1103/PhysRevResearch.7.L012081}{Phys. Rev. Research \textbf{7}, L012081 (2025).}






\bibitem{sup1} R. Gro\ss mann \textit{et al.}, Non-Gaussian Displacements in Active Transport on a Carpet of Motile Cells, \href{https://doi.org/10.1103/PhysRevLett.132.088301}{Phys. Rev. Lett. \textbf{132}, 088301 (2024).}

\bibitem{ss22} M. O. Costa, R. Silva, and D. H. A. L. Anselmo, Superstatistical and DNA sequence coding of the human genome, \href{https://doi.org/10.1103/PhysRevE.106.064407}{Phys. Rev. E \textbf{106}, 064407 (2022).}


\bibitem{traffic} A. Y. Abul-Magd, Modeling highway-traffic headway distributions using superstatistics, \href{https://doi.org/10.1103/PhysRevE.76.057101}{Phys. Rev. E \textbf{76}, 057101 (2007).}

\bibitem{Market0} E. Van der Straeten and C. Beck, Superstatistical fluctuations in time series: Applications to share-price dynamics and turbulence, \href{https://doi.org/10.1103/PhysRevE.80.036108}{Phys. Rev. E \textbf{80}, 036108 (2009).}



\bibitem{nature} B. Sch\"{a}fer, C. Beck, K. Aihara, D. Witthaut, and M. Timme, Non-Gaussian power grid frequency fluctuations characterized by Lévy-stable laws and superstatistics, \href{https://www.nature.com/articles/s41560-017-0058-z}{{Nat. Energy} \textbf{3}, 119 (2018).}


\bibitem{nature2} L. R. Gorjão \textit{et al.}, Open database analysis of scaling and spatio-temporal properties of power grid frequencies, \href{https://www.nature.com/articles/s41467-020-19732-7}{Nat. Communications \textbf{11}, 6362 (2020).}




\bibitem{ss4} G. Williams, B. Sch\"{a}fer, and C. Beck, Superstatistical
approach to air pollution statistics, \href{https://doi.org/10.1103/PhysRevResearch.2.013019}{Phys. Rev. Research \textbf{2}, 013019 (2020).} 






\bibitem{Plasma1} K. Ourabah, L. Aït Gougam, and M. Tribeche, Nonthermal and suprathermal distributions as a consequence of superstatistics, \href{https://doi.org/10.1103/PhysRevE.91.012133}{Phys. Rev. E \textbf{91}, 012133 (2015).}

\bibitem{Z4} F. Sattin and L. Salasnich, {
Multiparameter generalization of nonextensive statistical mechanics}, \href{https://doi.org/10.1103/PhysRevE.65.035106}{{{Phys. Rev.} E \textbf{65}, 035106(R) (2002).}}
\bibitem{splasma3} K. Ourabah, Demystifying the success of empirical
distributions in space plasmas, \href{https://doi.org/10.1103/PhysRevResearch.2.023121}{Phys. Rev. Research \textbf{2},
023121 (2020).} 




\bibitem{OurRelativistic} K. Ourabah, Superstatistics in the context of relativity, \href{https://doi.org/10.1103/7j5q-hxm2}{Phys. Rev. Research \textbf{7}, 033190 (2025).}

\bibitem{sup3} K. Ourabah, Fingerprints of nonequilibrium stationary distributions in dispersion relations, \href{https://doi.org/10.1038/s41598-021-91455-1}{Sci. Rep. \textbf{11}, 12103 (2021).}

\bibitem{Omar} O. Bouzit and K. Ourabah, Nonlinear structures in a nonequilibrium plasma: impact of small fluctuations, \href{https://doi.org/10.1140/epjp/s13360-024-04946-1}{Eur. Phys. J. Plus \textbf{139}, 175 (2024).}


\bibitem{Plasma2} S. Davis \textit{et al.}, Single-particle velocity distributions of collisionless, steady-state plasmas must follow superstatistics, \href{https://doi.org/10.1103/PhysRevE.100.023205}{Phys. Rev. E \textbf{100}, 023205 (2019).}
\bibitem{Plasma4} S. Davis \textit{et al.}, Kappa distribution from particle correlations in nonequilibrium, steady-state plasmas, \href{https://doi.org/10.1103/PhysRevE.108.065207}{Phys. Rev. E \textbf{108}, 065207 (2023).}




















\bibitem{Braginskii1965}
S. I. Braginskii, Transport processes in a plasma, 
Rev. Plasma Phys. \textbf{1}, 205 (1965).

\bibitem{Balescu1988}
R. Balescu,
\textit{Transport Processes in Plasmas, Vol. 1: Classical Transport Theory}
(North-Holland, Amsterdam, 1988).

\bibitem{Dum1990}
C. T. Dum,
in \textit{Physical Processes in Hot Cosmic Plasmas},
edited by W. Brinkmann, A.~C. Fabian, and F. Giovanelli
(Kluwer Academic, Dordrecht/Boston/London, 1990), pp. 157--180.





\bibitem{Hagelaar2005}
G. J. M. Hagelaar and L. C. Pitchford, Solving the Boltzmann equation to obtain electron transport coefficients and rate coefficients for fluid models, Plasma Sources Sci. Technol. \href{https://doi.org/10.1088/0963-0252/14/4/011}{\textbf{14}, 722 (2005).} 

\bibitem{Lv} C. Lv \textit{et al.}, Electronic transport of Lorentz plasma with collision and
magnetic field effects, \href{https://doi.org/10.1088/1674-1056/25/10/105201}{Chin. Phys. B \textbf{25}, 105201 (2016).} 

\bibitem{Du2013} J. Du, Transport coefficients in Lorentz plasmas with the power-law kappa-distribution, \href{https://doi.org/10.1063/1.4820799}{Phys. Plasmas \textbf{20}, 092901 (2013).}

\bibitem{Wang2017}
L. Wang and J. Du, The diffusion of charged particles in the weakly ionized plasma with power-law kappa-distributions, \href{https://doi.org/10.1063/1.4996775}{Phys. Plasmas \textbf{24}, 102305 (2017).}


\bibitem{Guo2019} R. Guo and J. Du, Transport coefficients of the fully ionized plasma with kappa-distribution and in strong magnetic field, \href{https://doi.org/10.1016/j.physa.2019.02.011}{Physica A \textbf{523}, 156 (2019).}



\bibitem{Husidic2021} E. Husidic, M. Lazar, H. Fichtner, K. Scherer, and S. Poedts, Transport coefficients enhanced by suprathermal particles in
nonequilibrium heliospheric plasmas, \href{https://doi.org/10.1051/0004-6361/202141760}{Astron. Astrophys. \textbf{654}, A99 (2021).}

\bibitem{Husidic2022} E. Husidic, K. Scherer, M. Lazar, H. Fichtner, and S. Poedts, Toward a Realistic Evaluation of Transport Coefficients in Non-equilibrium Space Plasmas, \href{https://doi.org/10.3847/1538-4357/ac4af4}{Astrophys. J. \textbf{927}, 159 (2022).}


\bibitem{Treumann} R. A. Treumann, Kinetic Theoretical Foundation of Lorentzian Statistical Mechanics, \href{https://doi.org/110.1238/Physica.Regular.059a00019}{Phys. Scr. \textbf{59}, 19 (1999).}









\bibitem{CLT} C. Beck, E. G. D. Cohen, and H. L. Swinney, From time series
to superstatistics, \href{https://doi.org/10.1103/PhysRevE.72.056133}{Phys. Rev. E \textbf{72}, 056133 (2005).} 






\bibitem{plasma1} B. Liu and J. Goree, Superdiffusion and Non-Gaussian Statistics in a Driven-Dissipative 2D Dusty Plasma, \href{https://doi.org/10.1103/PhysRevLett.100.055003}{Phys. Rev. Lett. \textbf{100}, 055003 (2008).}
\bibitem{Mahsa} M. Iranmanesh, H. Arjomand Kermani, and K. Ourabah, Superstatistics and stellar rotation: Modeling velocity distributions in six stellar groups, \href{https://doi.org/10.1016/j.newast.2025.102408}{New Astronomy \textbf{119}, 102408 (2025).}
\bibitem{VG} C. Brito, I. S. Aranson, and H. Chaté, {Vortex Glass and Vortex Liquid in Oscillatory Media}, \href{https://doi.org/10.1103/PhysRevLett.90.068301}{{{Phys. Rev. Lett.} \textbf{90}, 068301 (2003).}} 
\bibitem{invchi11} P. D. Dixit, A maximum entropy thermodynamics of small systems, \href{https://doi.org/10.1063/1.4804549}{J. Chem. Phys. \textbf{138}, 184111 (2013).}
\bibitem{Xu} D. Xu and C. Beck, Transition from lognormal to 
$\chi^2$-superstatistics for financial time series, \href{https://doi.org/10.1016/j.physa.2016.02.057}{Physica A \textbf{453}, 173 (2016).}


\bibitem{Jizba2} P. Jizba, J. Korbel, H. Lavička, M. Prokš, V. Svoboda, and C. Beck, Transitions between superstatistical regimes: Validity, breakdown and applications, \href{https://doi.org/10.1016/j.physa.2017.09.109}{Physica A \textbf{493}, 29 (2018).}
\bibitem{Feldman} W. C. Feldman, J. R. Asbridge, S. J. Bame, M. D. Montgomery, S. P. Gary, Solar wind electrons, \href{https://doi.org/10.1029/JA080i031p04181}{J. Geophys. Res. \textbf{80}, 4181 (1975).}

\bibitem{Salem} C. S. Salem, M. Pulupa, S. D. Bale, and D. Verscharen, Precision electron measurements in the solar wind at 1 au from NASA’s Wind spacecraft, \href{https://doi.org/10.1051/0004-6361/202141816}{Astron. Astrophys. \textbf{ 675}, A162 (2023).}
\bibitem{BGK} P. L. Bhatnagar, E. P. Gross, and M. Krook, A model for collision processes in gases. I. Small amplitude processes in charged and neutral one-component systems, \href{https://doi.org/10.1103/PhysRev.94.511}{Phys. Rev. \textbf{94}, 511 (1954).}

\bibitem{HelanderSigmar2005}
P. Helander and D. J. Sigmar, \textit{Collisional Transport in Magnetized Plasmas}
(Cambridge University Press, Cambridge, 2005).
\bibitem{BoydSanderson2003}
T. J. M. Boyd and J. J. Sanderson, \textit{The Physics of Plasmas} (Cambridge University Press, Cambridge, 2003).

\bibitem{Wang2018} Y. Wang and J. Du, The viscosity of charged particles in the weakly ionized plasma with power-law distributions, \href{https://doi.org/10.1063/1.5023030}{Phys. Plasmas \textbf{25}, 062309 (2018).}
\bibitem{Hagelaar2007} G. J. M. Hagelaar, Modelling electron transport in magnetized low-temperature discharge plasmas, \href{https://doi.org/10.1088/0963-0252/16/1/S06}{Plasma Sources Sci. Technol. \textbf{16}, S57 (2007).}


\bibitem{Risken1996}
H. Risken, \textit{The Fokker--Planck Equation: Methods of Solution and Applications}, 2nd ed. (Springer, Heidelberg, 1996).

\bibitem{operator} P. Andries, P. Le Tallec, J.-Ph. Perlat, and B. Perthame,
The Gaussian-BGK model of Boltzmann equation with small
Prandtl number, \href{https://doi.org/10.1016/S0997-7546(00)01103-1}{Eur. J. Mech. B Fluids \textbf{19}, 813 (2000).} 

\bibitem{Ourabahbook} K. Ourabah, \textit{Collective Phenomena in Plasmas and Elsewhere:Kinetic and Hydrodynamic Approaches} (Wiley-ISTE, Hoboken, 2023).

\bibitem{OurabahPRE2024} K. Ourabah, Superstatistics from a dynamical perspective: Entropy and relaxation, \href{https://doi.org/10.1103/PhysRevE.109.014127}{Phys. Rev. E \textbf{109}, 014127 (2024).}
\end{thebibliography}
\end{document}